\documentclass[fleqn,usenatbib]{mnras}

\usepackage{newtxtext,newtxmath}

\usepackage[T1]{fontenc}

\DeclareRobustCommand{\VAN}[3]{#2}
\let\VANthebibliography\thebibliography
\def\thebibliography{\DeclareRobustCommand{\VAN}[3]{##3}\VANthebibliography}

\usepackage{graphicx}	
\usepackage{amsmath}	

\usepackage[referable]{threeparttablex} 
\usepackage{booktabs} 
\usepackage{pdflscape}	
\usepackage{tabularx}
\usepackage{booktabs}
\usepackage{longtable} 
\usepackage{nccmath} 
\usepackage{xcolor}
\usepackage{ulem}

\newcommand{\nodata}{ ~$\cdots$~ } 

\title[Emergence and Sedimentation of Silicate Clouds in L Dwarfs]{Ultracool Dwarfs Observed with the \textit{Spitzer} Infrared Spectrograph. II. Emergence and Sedimentation of Silicate Clouds in L Dwarfs, and Analysis of the Full M5--T9 Field Dwarf Spectroscopic Sample} 

\author[Su\'arez \& Metchev]{
Genaro Su\'arez,$^{1}$\thanks{E-mail: gsuarez@uwo.ca}
Stanimir Metchev,$^{1,2}$
\\
$^{1}$Department of Physics and Astronomy, The University of Western Ontario, 1151 Richmond St, London, Ontario, N6A 3K7, Canada\\
$^{2}$Institute for Earth and Space Exploration, The University of Western Ontario, 1151 Richmond St, London, Ontario, N6A 3K7, Canada\\
}

\date{Accepted XXX. Received YYY; in original form ZZZ}

\pubyear{2015}

\begin{document}
\label{firstpage}
\pagerange{\pageref{firstpage}--\pageref{lastpage}}
\maketitle

\begin{abstract}
We present a uniform analysis of all mid-infrared $R\approx90$ spectra of field M5--T9 dwarfs obtained with the \textit{Spitzer} Infrared Spectrograph (IRS). The sample contains 113 spectra out of which 12 belong to late-M dwarfs, 69 to L dwarfs, and 32 to T dwarfs. Sixty-eight of these spectra are presented for the first time. We measure strengths of the main absorption bands in the IRS spectra, namely H$_2$O at 6.25~$\micron$, CH$_4$ at 7.65~$\micron$, NH$_3$ at 10.5~$\micron$, and silicates over 8--11~$\micron$. Water absorption is present in all spectra and strengthens with spectral type. The onset of methane and ammonia occurs at the L8 and T2.5 types, respectively, although ammonia can be detectable as early as T1.5. Silicate absorption sets in at spectral type L2, is on average the strongest in L4--L6 dwarfs, and disappears past L8. However, silicate absorption can also be absent from the spectra at any L subtype. We find a positive correlation between the silicate absorption strength and the excess (deviation from median) near-infrared colour at a given L subtype, which supports the idea that variations of silicate cloud thickness produce the observed colour scatter in L dwarfs. We also find that variable L3--L7 dwarfs are twice more likely to have above-average silicate absorption than non-variables. The ensemble of results solidifies the evidence for silicate condensate clouds in the atmospheres of L dwarfs, and for the first time observationally establishes their emergence and sedimentation between effective temperatures of $\approx$2000~K and $\approx$1300~K, respectively.
\end{abstract}



\section{Introduction}
\label{sec:intro}

Mid-infrared wavelengths ($\sim$5--50 $\micron$) sample prominent absorption features in the spectra of ultracool dwarfs, including water, methane, ammonia, and silicates \citep[e.g.][]{Roellig_etal2004,Cushing_etal2006,Looper_etal2008b}, as predicted by diverse models \citep[e.g.][]{Burrows_etal1997,Allard_etal2001,Marley_etal2002,Saumon_etal2003}. Prior to the \textit{Spitzer} Space Telescope \citep{Werner_etal2004}, these wavelengths remained spectroscopically unexplored in ultracool atmospheres. The Infrared Spectrograph \citep[IRS;][]{Houck_etal2004} on \textit{Spitzer} produced the first mid-infrared spectra of M, L, and T dwarfs \citep{Roellig_etal2004}, and revealed prominent absorption bands of H$_2$O at 6.25~$\micron$, CH$_4$ at 7.65~$\micron$, and NH$_3$ at 10.5~$\micron$. \citet{Cushing_etal2006} and \citet{Looper_etal2008b} further noted the presence of strong silicate absorption between 9--11~$\micron$ in the spectra of four mid-L dwarfs. Subsequent studies have expanded the mid-infrared spectroscopic sample, for a total of 243 published spectra of $\geq$M5 dwarfs, of which 77 are field dwarfs \citep[][]{Roellig_etal2004,Cushing_etal2006,Saumon_etal2007,Looper_etal2008b,Burgasser_etal2008,Leggett_etal2009,Stephens_etal2009,Kirkpatrick_etal2010,Leggett_etal2010b,Filippazzo_etal2015,Suarez_etal2021a} and 166 are members of young ($\lesssim$10~Myr) stellar associations, namely Chamaeleon I, Coronet, $\eta$ Chamaeleontis, Lupus, Ophiuchus, Serpens, Taurus, TW Hydra, and Upper Scorpius \citep[][]{Forrest_etal2004,Apai_etal2005,Furlan_etal2005,Furlan_etal2006,Bouwman_etal2006,Scholz_etal2007,Luhman_etal2007b,Merin_etal2007,Morrow_etal2008,Bouy_etal2008,Sicilia-Aguilar_etal2008,Riaz-Gizis2008,Furlan_etal2008,Pascucci_etal2009,Olofsson_etal2009,Riaz2009,Riaz_etal2009,McClure_etal2010,Manoj_etal2011,Furlan_etal2011,Adame_etal2011,Rilinger-Espaillat2021}. 

However, the published sample of field dwarfs observed with the {\it Spitzer} IRS comprises only 60 per cent of all field $\ge$M5 dwarfs with IRS observations, based on comparisons of existing compilations of ultracool dwarfs (Section \ref{sec:data_search}) with the \textit{Spitzer} Heritage Archive (SHA)\footnote{\url{https://sha.ipac.caltech.edu/applications/Spitzer/SHA/}}. In preparation for upcoming observations with the \textit{James Webb Space Telescope}, we have uniformly reprocessed and analysed all \textit{Spitzer} IRS observations of field ultracool dwarfs.

Herein we present all 113 field M5--T9 dwarfs that have detectable signal in their \textit{Spitzer} IRS low-resolution spectroscopic observations. All IRS observations were obtained during the \textit{Spitzer} cryogenic mission, which ended in 2009, before the discovery of the first Y dwarfs \citep{Cushing_etal2011}. None of the observed late-T dwarfs have since been re-classified as Y dwarfs. Hence, the presented compendium includes all field $\ge$M5 dwarfs with low-resolution \textit{Spitzer} IRS observations.
A subsequent publication will focus on young ultracool dwarfs with IRS spectra. In Section \ref{sec:data} we describe the observations and explain our updated procedure to extract the spectra. In Section \ref{sec:spectral_features} we analyse the strengths of the three main molecular species (H$_2$O, CH$_4$, and NH$_3$) and of 8--11~$\micron$ silicate absorption to infer dependencies on spectral type, near-infrared colours, photometric variability, or binarity. We summarise our work and conclude in Section \ref{sec:summary_conclusions}.

\section{Sample and Data Reduction}
\label{sec:data}

\subsection{The \textit{Spitzer} IRS Sample of Field M5--T9 Dwarfs}
\label{sec:data_search}
To search for spectroscopic observations of ultracool dwarfs obtained with the \textit{Spitzer} IRS, we queried the SHA on a comprehensive list of field late-M, L and T dwarfs. The list is based on the compilation by W.~Best and collaborators (The UltracoolSheet\footnote{\url{https://doi.org/10.5281/zenodo.4570814}} as of 2021 March 7), which contains about 3000 spectroscopically confirmed objects, mainly L and T brown dwarfs in the solar neighbourhood ($\lesssim$100~pc). The list was complemented with spectroscopically confirmed M5--M9 dwarfs from the compilation by J.~Gagne\footnote{\url{https://jgagneastro.com/list-of-m6-m9-dwarfs/}}, that contains over 8500 objects with such spectral types, most of them (about 60 per cent) M6 dwarfs. The final compiled catalogue has more than 11000 M5--M9, L and T dwarfs, the vast majority of them field objects.

We retrieved IRS short-low (SL) module spectroscopic observations for a total of 156 M5--T9 dwarfs from the compiled catalogue. We used a 60$\arcsec$ search radius and verified the matches by comparing the object names and by checking the sky coverage map in the SHA. Most observations were obtained in the standard staring mode. Four dwarfs (2MASS J00361617+1821104, 2MASS J02550357$-$4700509, 2MASS J05591914$-$1404488, and $\epsilon$~Ind~B) were observed in both staring mode and mapping mode, and two targets (2MASS J01075242+0041563 and 2MASS J21392676+0220226) were observed only in mapping mode. The vast majority of the observations include both SL orders (SL1 and SL2); only 12 targets lack SL2 module spectra. Thirty-one of the sources with IRS SL observations were also observed with the long-low (LL) spectroscopic module, 29 of which in both orders (LL1 and LL2). The SL and LL modules cover 5.2--14.8~$\micron$ and 13.9--38~$\micron$, respectively, at $R\approx90$ in order 2 of both modules.

Thirteen of the M5--T9 dwarfs with low-resolution IRS data (2MASS names; J00361617+1821104, J01365662+0933473, J01390120$-$1757026, J02550357$-$4700509, J04234858$-$0414035, J05591914$-$1404488, J08294949+2646348, J12255432$-$2739466, J12545393$-$0122474, J14563831$-$2809473, J19165762+0509021, J22041052$-$5646577, and J22383372$-$1517573) were also observed at high resolution with the IRS, six of which were published in \citet{Mainzer_etal2007}. However, there are no additional $\geq$M5 dwarfs with only high-resolution observations. In this work we present only the low-resolution IRS data.

Of the 156 M5--T9 dwarfs with \textit{Spitzer}/IRS SL observations in our compiled catalogue, 121 are field objects and 35 are known members of young ($\lesssim$10~Myr) stellar associations, namely Chamaeleon I, Lupus, Ophiuchus, Taurus, TW Hydra, and Upper Scorpius. Our present catalogue is thus very incomplete to IRS observations of $\geq$M5 dwarfs in such young stellar associations, of which we found 166 in the published literature (Section~\ref{sec:intro}). In the present paper we focus on the field dwarfs, and defer the discussion of $\lesssim$10~Myr old $\geq$M5 dwarfs to a follow-up paper. We note that while the vast majority of field dwarfs in our sample are old ($\gtrsim$1~Gyr) objects, there are also some younger dwarfs. One set of examples includes objects that exhibit spectroscopic signatures of youth, such as: 0501$-$0010 and 1022+5825 \citep[with very low and intermediate surface gravity, respectively;][]{Cruz_etal2009}, the $\approx$300 Myr old dwarf HN~Peg~B \citep[a.k.a., 2144+1446;][]{Luhman_etal2007,Suarez_etal2021a}, and the 40--400~Myr old dwarf 0758+3247 \citep{Stephens_etal2009}. Others include objects that are kinematic members of nearby young moving groups, for instance: 0355+1133 and 1425$-$3650 (AB Doradus, $\sim$100~Myr; \citealt{Liu_etal2013,Gagne_etal2015}), and 0136+0933 and 2139+0220 (Carina-Near, $\sim$200~Myr; \citealt{Gagne_etal2017,Zhang_etal2021}). At such moderately young (few tens to hundreds of Myr) ages, these dwarfs are not expected to have mid-infrared excesses caused by circumstellar disks \citep[e.g.;][]{Calvet_etal2005,Briceno_etal2005,Downes_etal2014}. We have visually verified this in their IRS spectra (Figure~\ref{fig:spectra_SL_1}--\ref{fig:spectra_SL_3}). Therefore, we include both moderately young (few tens to hundreds of Myr) and old ($\gtrsim$1~Gyr) field dwarfs in the following analysis and, for simplicity, collectively refer to them as `field' dwarfs.

As we detail in Section~\ref{sec:data_reduced}, 113 of the 121 field $\geq$M5 dwarfs targeted with the IRS have usable spectra. Tables \ref{tab:param} and \ref{tab:log} list, respectively, select parameters (name, coordinates, discovery paper, spectral type, binarity, and variability) and the observing log (\textit{Spitzer} AOR key, program ID and PI, spectroscopic module, exposure time, and observation date) of the 113 field dwarfs for which we present IRS low-resolution spectra. Hereafter we abbreviate the names of the targets as hhmm$\pm$ddmm, where the suffix is the right ascension (hours and minutes) and declination (degrees and minutes) at the J2000.0 equinox. We list optical and infrared spectral types (when available) of the objects, and prioritise optical spectral types for M and L dwarfs and infrared types for T dwarfs. Table \ref{tab:log_nullSNR} lists the 8 field $\geq$M5 dwarfs that were targeted with the IRS, but for which the data did not yield good extractions. These spectra have S/N$\sim$0 because either the exposure times were too short or the targets were off the slit. 

\subsection{Improved Reduction of the \textit{Spitzer} IRS SL and LL Observations}
We downloaded from the SHA all basic calibrated data (BCD)---also known as Level 1 products---of the targets in the input list with IRS low-resolution observations. The BCD are produced by the IRS pipeline after ramp fitting, dark subtraction, droop correction, linearity correction, flat fielding, and wavelength calibration. We further reduced the data by removing bad pixels (outliers and rogues) from the 2D spectral images using the IRSCLEAN tool\footnote{\url{https://irsa.ipac.caltech.edu/data/SPITZER/docs/dataanalysistools/tools/irsclean/}} with the appropriate IRS campaign mask and with different \texttt{noise\_floor} and \texttt{aggressive} parameters (ranging between 0.1 and 0.5, and 1 and 2, respectively) for each target to remove as many bad pixels as possible without affecting the spectrum trace. We combined the 2D spectral images taken at the same nod position (for staring mode data) or map position (for mapping mode data) using the \textit{coad} script\footnote{\url{https://irsa.ipac.caltech.edu/data/SPITZER/docs/dataanalysistools/tools/coad/}} provided by the Spitzer Science Center (SSC). We then removed the background of the combined 2D spectral images in staring mode by pairwise subtracting them. Any residual background was further removed by subtracting from each image row (i.e., perpendicular to the trace) the median flux of that row after excluding both nod traces. To remove background from the 2D spectral images obtained in mapping mode, we subtracted a median-combined image with the target off the slit from a median-combined image with the target in the slit. Any residual background was further removed as for the staring mode data.

Before extracting the spectra, we again ran IRSCLEAN to remove residual bad pixels marked by hand on the background-subtracted images. The spectroscopic extraction was performed with the \textit{Spitzer} IRS Custom Extraction (SPICE) software\footnote{\url{https://irsa.ipac.caltech.edu/data/SPITZER/docs/dataanalysistools/tools/spice/}} developed by the SSC. We used the default trace widths (4 pixels at 6~$\mu$m, 8 pixels at 12~$\mu$m, 4.25 pixels at 16~$\mu$m, and 7.172 pixels at 27~$\mu$m for the SL2, SL1, LL2, and LL1 spectroscopic modules) to obtain flux-calibrated spectra (in Jy). We evaluate the accuracy of this flux calibration in Section~\ref{sec:calibration_accuracy}. Spectra obtained with the same module and in the same order were averaged. When assembling observations in multiple orders or modules, we used SL2 spectra for $5.2\le\lambda~(\micron)\le7.5$ and SL1 for longer wavelengths in the SL module (up to 14.1 $\micron$), and LL2 for $14.1<\lambda~(\micron)\le20.5$ and LL1 for $20.5<\lambda~(\micron)\le38$ in the LL module. Thus, for targets observed in both SL and LL modules, we present a single 5.2--38~$\mu$m spectrum. Any offsets between the flux densities among the modules were removed by scaling the LL spectrum to the SL spectrum in the 13.9--14.8~$\micron$ overlap region.

The uncertainties of the spectra were obtained in three steps. First, when combining the BCD, an uncertainty 2D image (\textit{c2unc}) was created considering the scatter of the individual frames. Then, in the background subtraction process new uncertainty frames (\textit{bkunc}) were created by adding in quadrature the \textit{c2unc} uncertainties of each pair of 2D images. Finally, the \textit{bkunc} uncertainties of the pixels at each wavelength resolution element in the extraction trace widths were added in quadrature to obtain the uncertainties of the extracted spectra.

\subsection{Reduced Low-resolution IRS Spectra for All 113 Field M5--T9 Dwarfs in the SHA}
\label{sec:data_reduced}
\begin{figure*}
	\centering
	\includegraphics[width=0.99\linewidth]{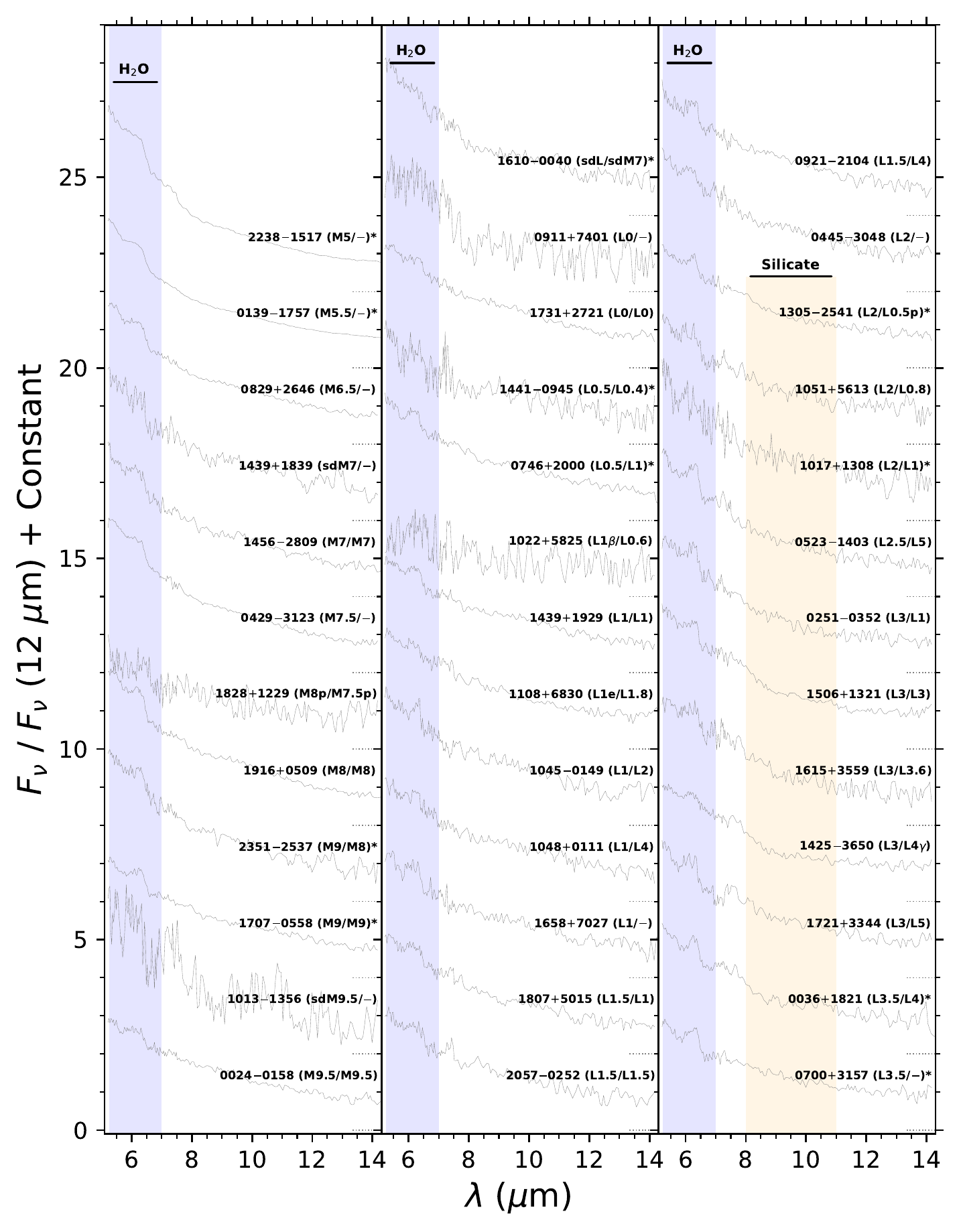}
	\caption[]{\textit{Spitzer} IRS 5.2--14.2 $\micron$ spectra of 12 M5--M9 dwarfs (left panel) and 26 L0--L3 dwarfs (middle and right panels). The spectra were normalised to unity using the median flux at 12~$\micron$ in a 0.6~$\micron$ window and offset by constants (dotted lines). Binary sources are indicated with asterisks (Table \ref{tab:param}). The optical and near-infrared spectral types (when available) of the dwarfs are indicated in the labels. The absorption regions for water and silicates are shaded. \\ (The spectra shown in this figure are available in the online journal.)}
	\label{fig:spectra_SL_1}
\end{figure*}

\begin{figure*}
	\centering
	\includegraphics[width=0.99\linewidth]{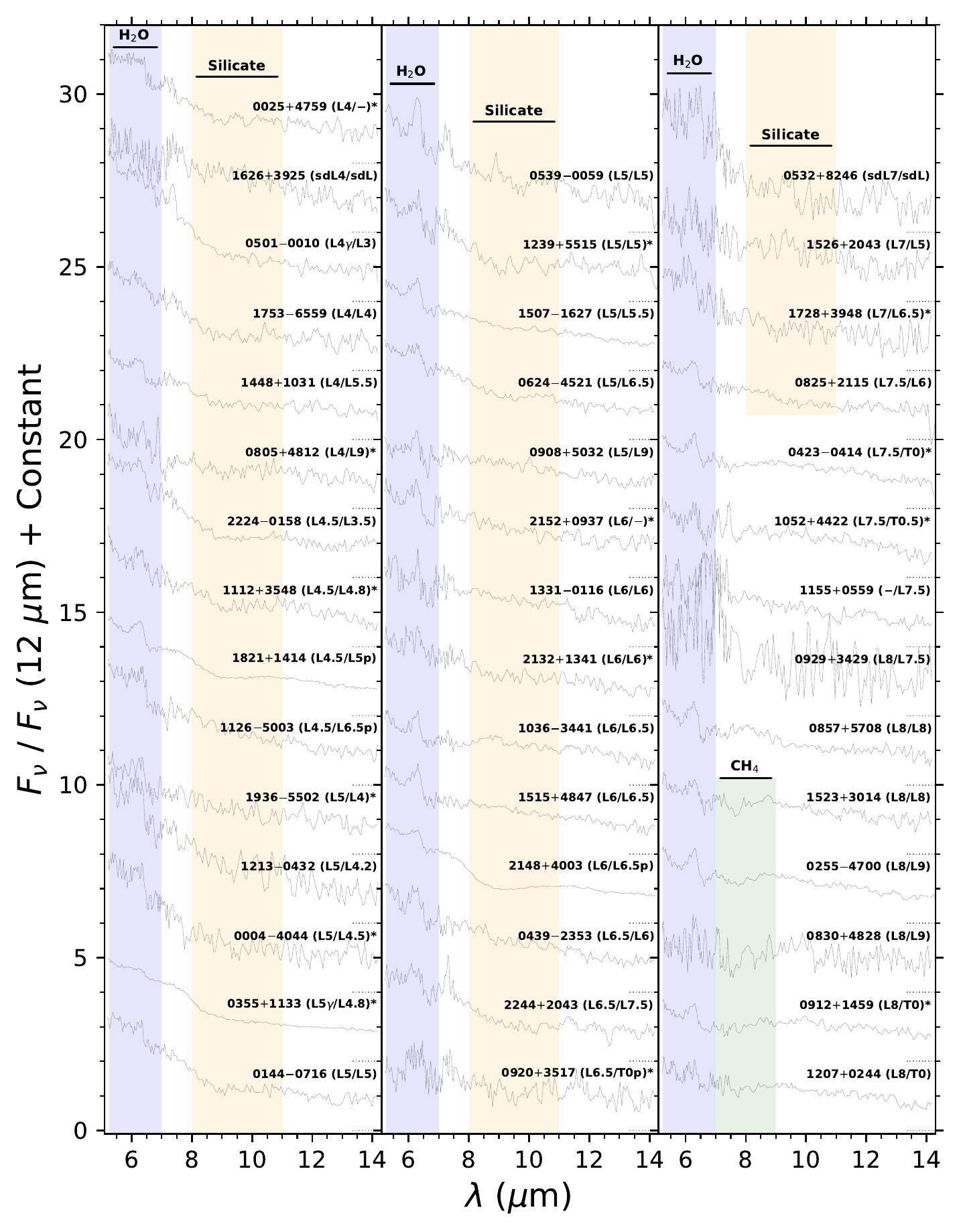}
	\caption[]{\textit{Spitzer} IRS 5.2--14.2 $\micron$ spectra of 43 L4--L8 dwarfs.The spectra were normalised to unity using the median flux at 12~$\micron$ in a 0.6~$\micron$ window and offset by constants (dotted lines). Binary sources are indicated with asterisks (Table \ref{tab:param}). The optical and near-infrared spectral types (when available) are indicated in the labels. The absorption regions for water, methane, and silicates are shaded. \\ (The spectra shown in this figure are available in the online journal.)}
	\label{fig:spectra_SL_2}
\end{figure*}

\begin{figure*}
	\centering
	\includegraphics[width=0.99\linewidth]{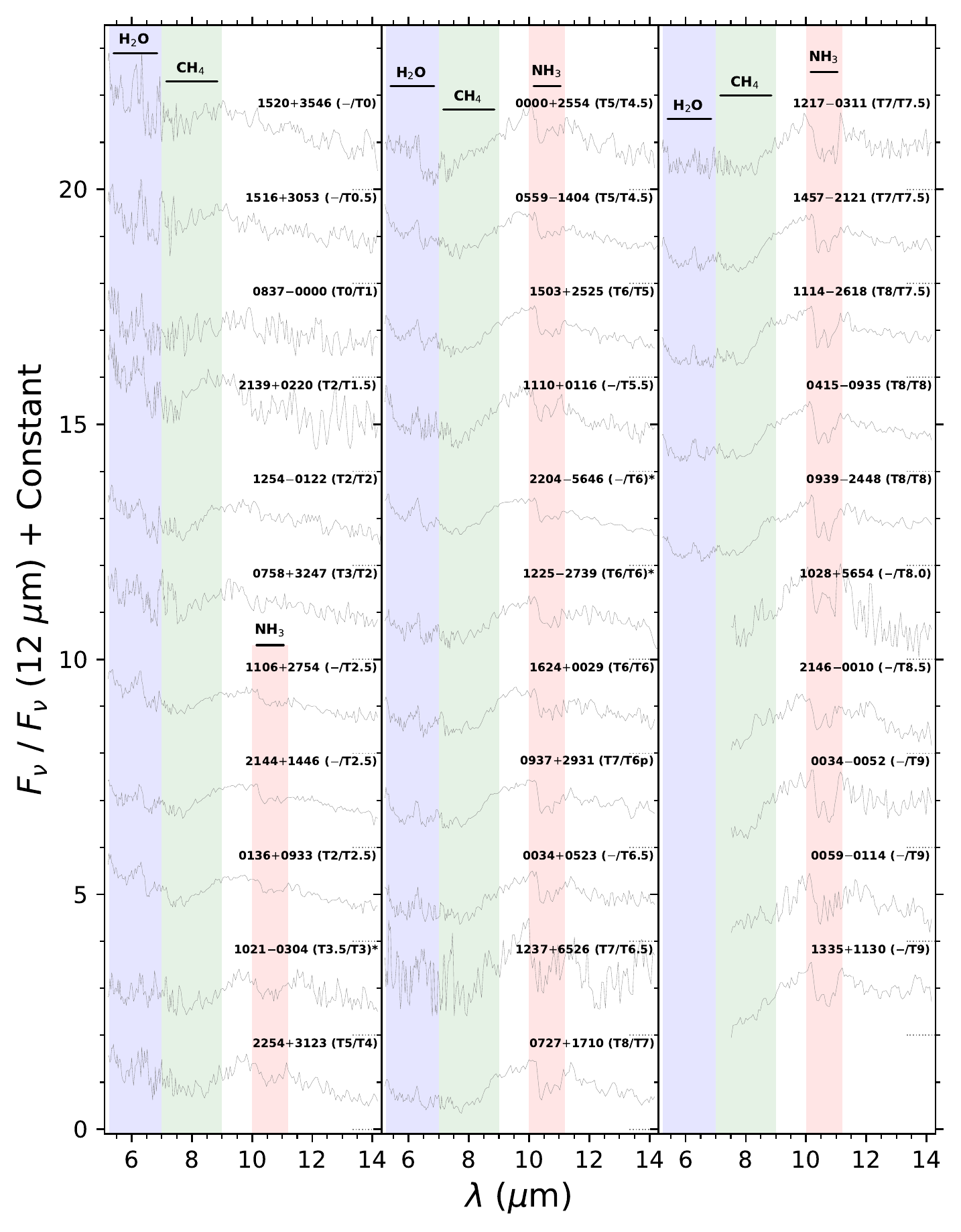}
	\caption[]{\textit{Spitzer} IRS  5.2--14.2~$\micron$ spectra of 27 T0--T8 dwarfs and 7.5--14.2~$\micron$ IRS spectra of 5 T8--T9 dwarfs. The spectra were normalised to unity using the median flux at 12~$\micron$ in a 0.6~$\micron$ window and offset by constants (dotted lines). Binary sources are indicated with asterisks (Table \ref{tab:param}). The optical and near-infrared spectral types (when available) are indicated in the labels. The absorption regions for water, methane, and ammonia are shaded. \\ (The spectra shown in this figure are available in the online journal.)}
	\label{fig:spectra_SL_3}
\end{figure*}

\begin{figure*}
	\centering
	\includegraphics[width=0.99\linewidth]{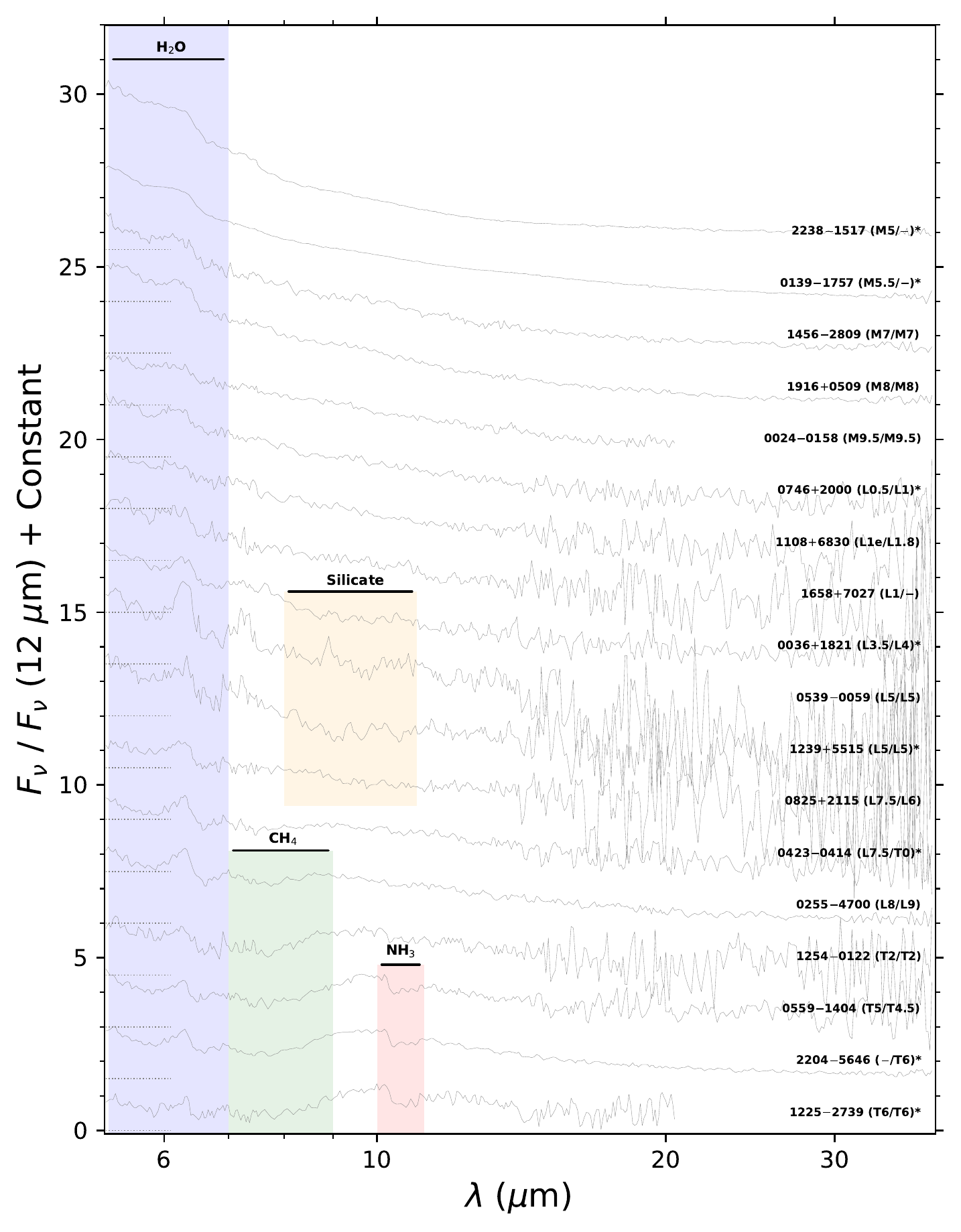}
	\caption[]{\textit{Spitzer} IRS 5.2--38 $\micron$ spectra of 18 M5--T6 dwarfs (5.2--20.5 $\micron$ for 0024$-$0158 and 1225$-$2739). The spectra were normalised to unity using the median flux at 12~$\micron$ in a 0.6~$\micron$ window and offset by constants (dotted lines). Binary sources are indicated with asterisks (Table \ref{tab:param}). The optical and near-infrared spectral types (when available) are indicated in the labels. The absorption regions for water, methane, silicates, and ammonia are shaded. \\ (The spectra shown in this figure are available in the online journal.)}
	\label{fig:spectra_SL_LL}
\end{figure*}

We successfully extracted 113 spectra of the 121 field M5--T9 dwarfs observed with the IRS. Twelve of these 113 spectra belong to M5--M9 dwarfs, 69 to L-type dwarfs, and 32 to T-type dwarfs. In Figures~\ref{fig:spectra_SL_1}, \ref{fig:spectra_SL_2}, and \ref{fig:spectra_SL_3} we show the 5.2--14.2~$\micron$ IRS spectra (7.5--14.2~$\micron$ for the five latest T dwarfs) of the objects with SL module observations. Eighteen of the 113 spectra have both SL and LL data. Figure~\ref{fig:spectra_SL_LL} shows the 5.2--38~$\micron$ IRS spectra (5.2--20.5~$\micron$ for 0024$-$0158 and 1225$-$2739) of these 18 targets. In Table~\ref{tab:log} we list all 113 dwarfs together with the signal-to-noise ratio (S/N) at 6 and 12~$\mu$m of the IRS spectra. The median S/N is $\approx$20 at 6~$\mu$m and $\approx$10 at 12~$\mu$m.

Our final sample contains \textit{Spitzer} IRS spectra of 36 M7--T9 dwarfs that are discussed for the first time. The remaining 77 M5--T9 dwarfs with IRS low-resolution spectra have already been discussed in preceding studies \citep{Roellig_etal2004,Cushing_etal2006,Saumon_etal2007,Looper_etal2008b,Burgasser_etal2008,Leggett_etal2009,Stephens_etal2009,Kirkpatrick_etal2010,Leggett_etal2010b,Filippazzo_etal2015,Suarez_etal2021a}, as referenced in Table \ref{tab:log}. However, spectra for 32 of these 77 previously analysed objects have not been shown or made publicly available. Therefore, our work makes IRS spectra for 68 ultracool dwarfs, over half of the field ultracool IRS sample, available for the first time.

\subsection{Accuracy of the IRS Flux Calibration}
\label{sec:calibration_accuracy}

We evaluated the accuracy of the automatic flux calibration of the IRS spectra by comparing IRAC channel 4 (8.0~$\micron$, [8.0]) and WISE W3 photometry of the targets with synthetic photometry from the IRS spectra. For the [8.0]-band comparisons we considered sources having both SL module orders (SL1 and SL2) to get full spectral coverage of the [8.0] pass band (6.30--9.59 $\micron$). On the other hand, to cover the entire W3 pass band (7.44--17.26 $\micron$), we require IRS spectra with both SL and LL modules (at least LL2 order). From our sample of 113 IRS spectra, 108 meet the first criterion and 18 meet the second criterion. We obtained the synthetic flux $F_{\rm syn}$ over the band pass ($\lambda_A, \lambda_B$) by convolving the filter response $S_{\lambda}$ with the synthetic spectrum $F_{\lambda}$:

\begin{ceqn}
\begin{align}
	F_{\rm syn} = \dfrac{\int_{\lambda_A}^{\lambda_B}F_{\lambda}S_{\lambda}d\lambda}{\int_{\lambda_A}^{\lambda_B}S_{\lambda}d\lambda}
	\label{eq:synt_phot}
\end{align}
\end{ceqn}

The synthetic fluxes were converted to synthetic magnitudes $m_{\rm syn}$ to be compared with the observed magnitudes $m_{\rm obs}$ listed in Table~\ref{tab:photometry}. For 67 of the 70 dwarfs with observed [8.0] magnitudes we calculated the corresponding [8.0]-band synthetic magnitudes, as they have full spectral coverage of the [8.0] filter pass band. Fourteen of these also had IRS LL2 coverage that enabled comparisons of their synthetic and observed W3 magnitudes. Four additional dwarfs without observed [8.0] magnitudes had \textit{WISE} W3 magnitudes, and also sufficient spectral coverage to calculate the corresponding synthetic W3 magnitudes. Thus, a total of 71 M5.5--T8 dwarfs had [8.0] or W3 photometry and sufficient IRS spectral coverage to compare measured vs.\ synthetic fluxes.

\begin{figure}
	\centering
	\includegraphics[width=1.0\linewidth]{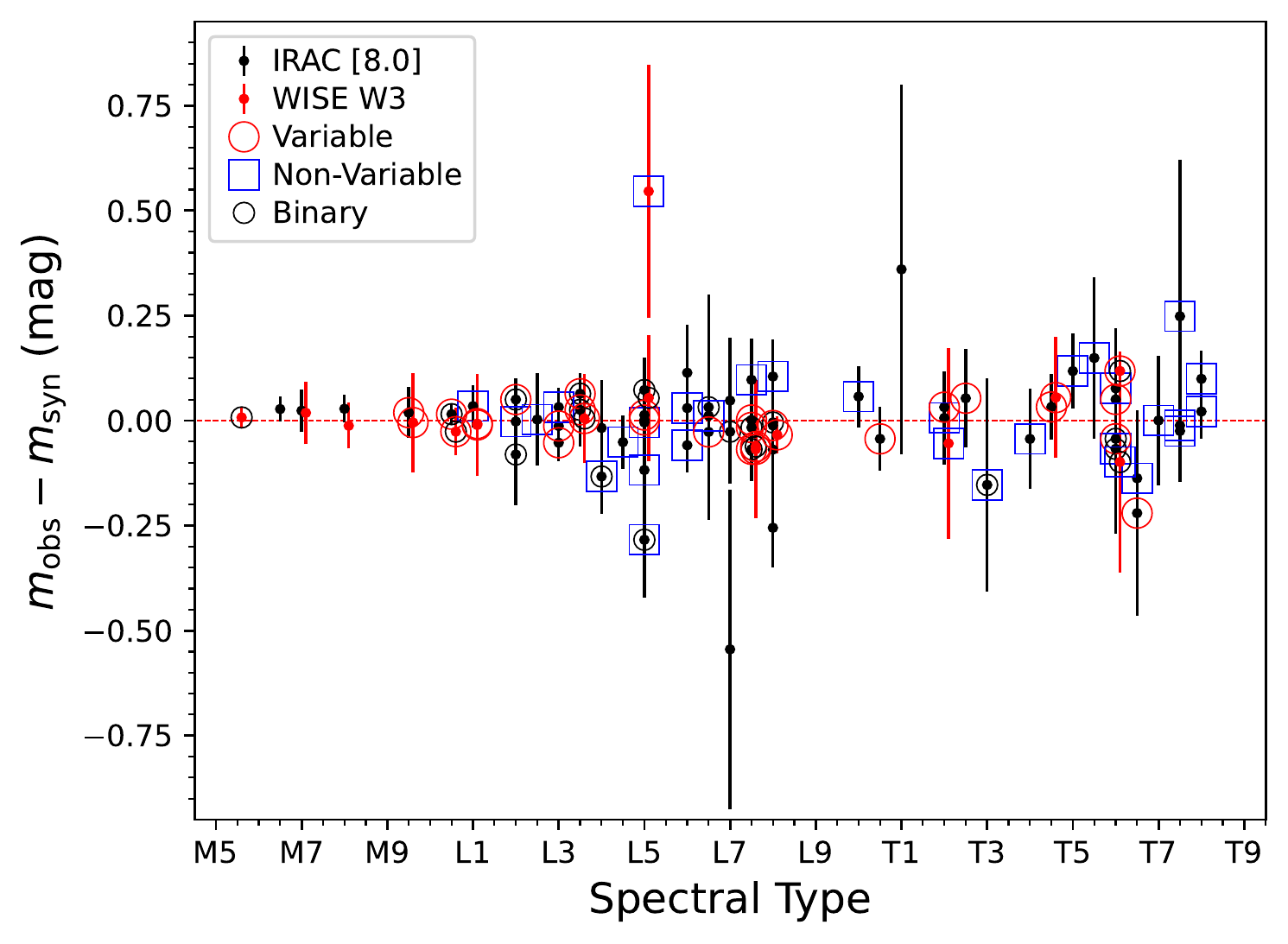}
	\caption{Residuals between observed and synthetic magnitudes using the IRAC [8.0] (black points) and WISE W3 (red points) pass bands as a function of spectral type after correcting by the average offset of $-0.064\pm0.014$ mag along the whole spectral type range (see Section~\ref{sec:calibration_accuracy}). The synthetic photometry was obtained from the IRS spectra presented in this study. The corrected 8 and 12~$\micron$ IRS fluxes scatter with an RMS of 14 per cent. The W3 comparison points are shifted slightly to the right for clarity. The red circles, blue squares, and black circles indicate variable, non-variable, and binary sources (Table \ref{tab:param}), respectively.}
	\label{fig:flux_calib}
\end{figure}

For the comparison we excluded three sources with differences between measured and synthetic magnitudes that were larger than 5 times the standard error of the mean (SEM; $\sigma/\sqrt{N}$, where $\sigma$ is the data scatter and $N$ is the sample size). Two are outliers (0000+2554 and 0501$-$0010) in the [8.0] photometric comparison and one (2238$-$1517) at W3. (In Section~\ref{sec:calibration_outliers} we further improve the flux calibration of these and other discrepant objects). After outlier removal, the average and SEM of the [8.0] and W3 residuals are mutually consistent: $-0.071\pm0.015$ mag and $-0.039\pm0.034$ mag, respectively. When considering both band passes together, the difference between the observed and synthetic magnitudes is $-0.064\pm0.014$ mag. We added this $-0.064$ mag average offset to our synthetic magnitudes reported in Table \ref{tab:photometry} and applied the corresponding scaling (by a factor of 1.06) to all IRS spectra presented in this study. In Figure \ref{fig:flux_calib} we show the residuals between observations and synthetic photometry after applying the flux calibration correction. 

\subsection{Improvement of the IRS Flux Calibration for Individual Targets}
\label{sec:calibration_outliers}

We evaluated in Section \ref{sec:calibration_accuracy} the overall IRS flux calibration accuracy of field M5--T9 dwarfs with full IRS wavelength coverage of either the IRAC [8.0] or the WISE W3 pass bands. However, we could not verify the automatic flux calibration for about 40 per cent of our 113 objects because they lack SL2 and/or LL2 observations, and so do not fully cover the [8.0] or W3 filter pass bands. In order to provide the community with a flux-calibrated mid-infrared spectroscopic sample of ultra-cool dwarfs, we further assess the flux calibration of the IRS spectra as follows.

The five latest (T8--T9) dwarfs in our sample were not observed in SL2 mode, and so lack spectra shortward of 7.50~$\micron$. To complement their wavelength coverage so that it fully encompasses the [8.0] 6.30--9.59~$\micron$ filter pass band, we extended the spectra from 7.50~$\micron$ to 6.30~$\micron$ using the average IRS SL2 spectrum of the T8 dwarfs 0415$-$0935 and 0939$-$2448 (Figure~\ref{fig:spectra_SL_3}).

A larger number (95) of our 113 targets were not observed in LL2 mode (Table~\ref{tab:log}), and so lack spectra longward of 14.2~$\micron$. To complement their wavelength coverage so that it fully encompasses the W3 7.44--17.26~$\micron$ filter pass band, we extended the spectra from 14.20~$\micron$ to 17.26~$\micron$ with the LL2 spectrum of an object most similar in spectral type (Figure \ref{fig:spectra_SL_LL}). When two templates with the same spectral type and similar S/N are available, we used the average IRS spectrum. The templates cover spectral types M5--T6 in steps of no more than 3 subtypes. Thus, each extension was done using a template within 1.5 subtypes from the spectral type of the target, except for the latest dwarfs, for which the template has a spectral type difference of up to 3 subtypes. However, as the IRS LL M5--T9 spectra appear featureless in the extension region (Figure \ref{fig:spectra_SL_LL}), the IRS spectra at these wavelengths have essentially the same Rayleigh-Jeans behaviour.

We used Equation~\ref{eq:synt_phot} to obtain synthetic [8.0] and W3 photometry from all extended IRS spectra scaled up by the overall factor of 1.06 determined in Section~\ref{sec:calibration_accuracy}. The resulting synthetic magnitudes were compared to the observed IRAC [8.0] and WISE W3 magnitudes (Table~\ref{tab:photometry}). If the synthetic and the observed magnitudes were more than 3$\sigma$ or 0.5 mag discrepant, we scaled the IRS spectrum to the observed photometry. When both [8.0] and W3 observed magnitudes were available for the same object, the IRS flux of the object was corrected only if the criterion was met in both comparisons.

We applied this additional calibration of the IRS fluxes to a total of 13 of our 113 M5--T9 dwarfs, including the three outliers mentioned in Section~\ref{sec:calibration_accuracy} (0000+2554, 0501$-$0010, and 2238$-$1517) and 10 additional dwarfs (0532+8246, 0830+4828, 0911+7401, 1022+5825, 1106+2754, 1213$-$0432, 1441$-$0945, 1821+1414, 2139+0220, and 2148+4003). Two of these 13 objects were identified from the [8.0] magnitude comparison, nine from the W3 comparison, and one from both comparisons.

The likely reason for the highly discrepant IRS vs.\ [8.0] and/or W3 fluxes for these 13 objects is inadequate positioning on the IRS slit because of inaccurate knowledge of the objects' precise coordinates. The M5--T9 dwarfs in the sample are generally too faint for on-target acquisition with the blue (15~$\micron$) or red (22~$\micron$) IRS peak-up imaging (PUI) camera, and so would have required offset acquisition with respect to a nearby $V<10$~mag target using the 505--595~nm Pointing Calibration and Reference Sensor (PCRS) on Spitzer \citep{Houck_etal2004}. Doing so requires accurate knowledge of the proper motion of the science target, which for our M5--T9 dwarfs can be significant ($\gtrsim200$~mas yr$^{-1}$). However, during the pre-WISE and pre-Gaia 2004--2009 epoch of Spitzer IRS observations, the proper motions would not have been known to sufficient accuracy to centre the targets on the Spitzer IRS slit (3.7$\arcsec$ width for SL1). Consequently, the lost flux would not have been accounted for in the SSC pipeline extraction, resulting in a significant discrepancy between the IRS fluxes and the [8.0] or W3 band photometry. We believe this to be the case for the most discrepant IRS spectra, given that a further eight of the 121 field M5--T9 dwarfs targeted with IRS did not yield any detectable signal at all (Table~\ref{tab:log_nullSNR}): as would be expected from even poorer alignment.

We used the fully flux-calibrated IRS spectra of the 113 field M5--T9 dwarfs to analyse their main mid-infrared spectral features in the next section.

\section{Spectral Features}
\label{sec:spectral_features}
\subsection{Molecules: Water, Methane, and Ammonia}
\label{sec:water_methane_ammonia}
We measured the depths of the H$_2$O, CH$_4$, and NH$_3$ absorption at 6.25 $\micron$, 7.65 $\micron$, and 10.5 $\micron$, respectively, considering the spectral index definitions in \citet{Cushing_etal2006}. We modified the \citet{Cushing_etal2006} definition for the methane index to sample the region of strongest absorption, at the 7.65~$\micron$ methane band head, instead of at 8.5~$\micron$:

\begin{ceqn}
\begin{align}
	{\rm H}_2{\rm O\ Index} = \frac{F_{6.25}}{0.562F_{5.80}+0.474F_{6.75}},
	\label{eq:H2O}
\end{align}
\end{ceqn}

\begin{ceqn}
\begin{align}
	{\rm CH}_4{\rm\ Index} = \frac{F_{10.0}}{F_{7.65}},
	\label{eq:CH4}
\end{align}
\end{ceqn}

\begin{ceqn}
\begin{align}
	{\rm NH}_3{\rm\ Index} = \frac{F_{10.0}}{F_{10.8}}.
	\label{eq:NH3}
\end{align}
\end{ceqn}
$F_\lambda$ above is the mean flux density around the specified wavelength (in $\micron$) in a window of 0.3 $\micron$ for the water index and 0.6 $\micron$ for both the methane and the ammonia indices. For an illustration of these index definitions we refer the reader to Figure 6 in \citet{Cushing_etal2006}.

\begin{figure}
	\centering
	\includegraphics[width=1.0\linewidth]{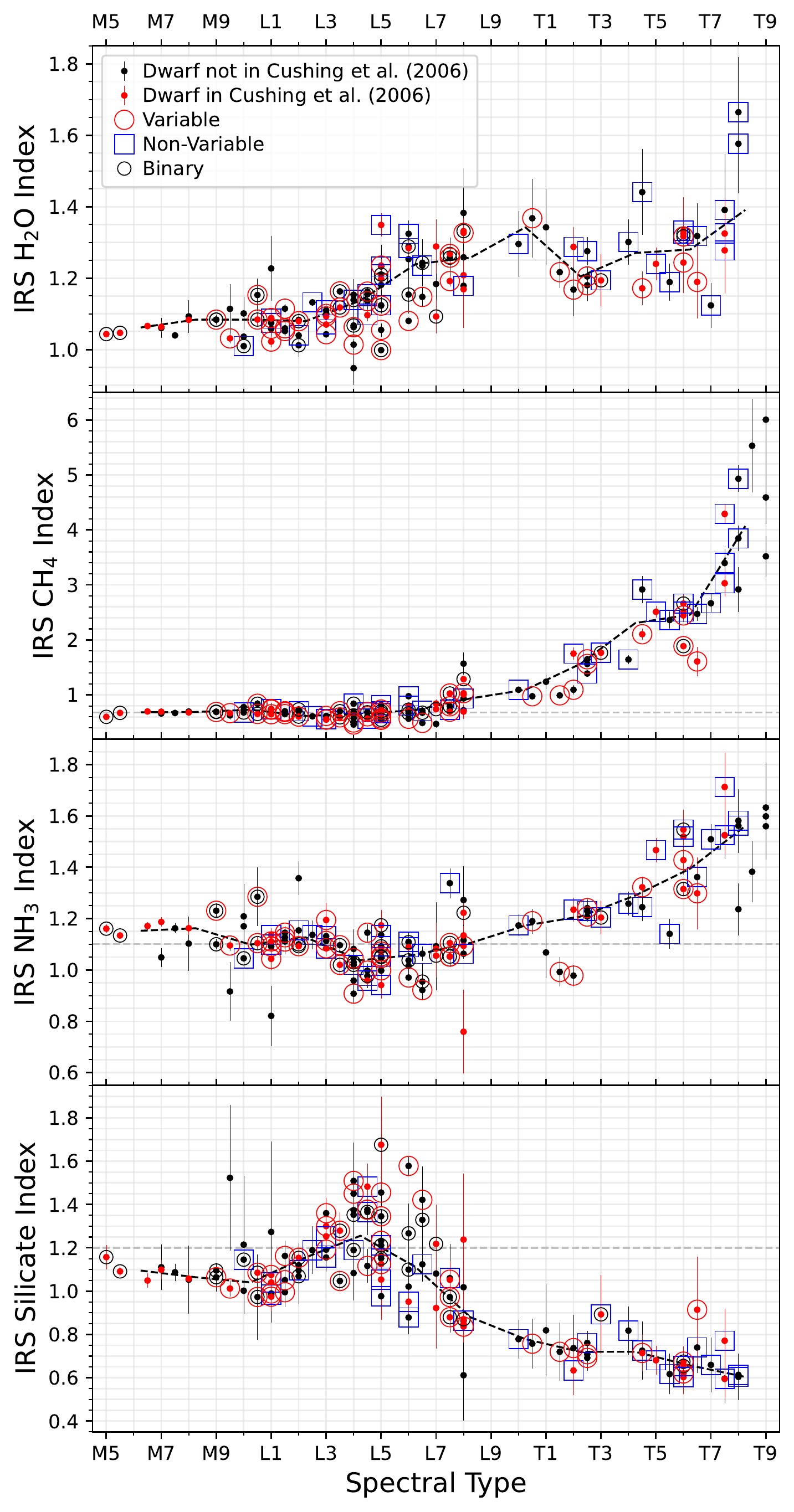}
	\caption{The H$_2$O, CH$_4$, NH$_3$, and silicate indices from the IRS spectra as a function of spectral type. The dwarfs included in the \citet{Cushing_etal2006} list are shown with red points; the dwarfs absent from that reference are shown with black points. The red circles, blue squares, and black circles indicate variable, non-variable, and binary sources (Table \ref{tab:param}), respectively. The dashed black curve in each panel traces the median spectral index in bins of two spectral subtypes. The dashed grey line in the middle panels is a reference for the onset of the methane and ammonia absorption bands and is obtained as the median index value for $\le$L5-type (where the features are clearly absent in the IRS spectra) objects. The dashed grey line in the bottom panel shows the threshold above which we selected sources with the largest silicate index values.}
	\label{fig:indices}
\end{figure}

We computed the above three spectral indices for most dwarfs in our sample. The water index for the five latest T dwarfs could not be computed because, as they lack $\lambda \leq 7.5~\micron$ IRS SL2 data (Figure \ref{fig:spectra_SL_3}; Table \ref{tab:log}), there is no full spectral coverage of the absorption. In Table~\ref{tab:indices} we list the index measurements of the main molecular features (H$_2$O, CH$_4$, and NH$_3$) in the IRS spectra of 113 field M5--T9 dwarfs. The index uncertainties were obtained from the uncertainties in the mean flux densities of both the absorption and the continuum. In the first three panels of Figure \ref{fig:indices} we show the index values as a function of spectral type. The median index values in spectral type bins of two subtypes (to average down statistical scatter) are connected with a dashed black line for each absorber. We highlighted with a different colour (red points) the targets in the \citet{Cushing_etal2006} sample to distinguish them from the new dwarfs in our sample (black points). We increased the sample of M5--T9 dwarfs with IRS molecular absorption measurements from 38 to 113 objects (a $\approx$200 per cent increase). We notice the significant addition of dwarfs particularly at early-T types.

We see from Figure \ref{fig:indices} that H$_2$O absorption overall strengthens with spectral type from M5 to T9, even if it remains roughly constant within a large scatter between L8--T7. The onset of CH$_4$ absorption at the resolution and S/N of the IRS data is at about spectral type L8, and rapidly strengthens toward later spectral types. The actual spectral type of CH$_4$ onset is subject to ambiguity in the optical vs.\ near-infrared spectral classification or binarity of L/T-transition dwarfs. In particular, the unresolved binary 0423$-$0414, which has a composite spectral type of L7.5/T0 (optical/near-infrared) and for which we adopt a spectral type of L7.5 in our analysis, already exhibits CH$_4$ absorption at 7.65~$\micron$ (Figures~\ref{fig:spectra_SL_2} and \ref{fig:spectra_SL_LL}). The individual components of 0423$-$0414 have spectral types of L$6\pm1$ and T$2\pm1$ \citep{Burgasser_etal2005}. Hence, the CH$_4$ absorption is likely caused by the T2 component.

We also see in Figure \ref{fig:indices} that NH$_3$ absorption becomes detectable first in 2139+0220 at (near-infrared) spectral type of T1.5 and is consistently present in $\ge$T2.5 dwarfs (Figure~\ref{fig:spectra_SL_3}). Ammonia strengths increase marginally toward cooler temperatures.

Overall, our assessment of mid-infrared molecular absorbers is consistent with the findings in \citet{Cushing_etal2006}, who report that both CH$_4$ and NH$_3$ set in around the L/T transition in \textit{Spitzer} IRS spectra. Our uniform analysis of the more comprehensive sample increases the robustness of these results.

Known binaries and variables scatter around the median index values of the water, methane, and ammonia absorption bands (top three panels of Figure~\ref{fig:indices}). That is, neither the composite spectroscopic signature of an unresolved binary nor the presence of photometric variability discernibly bias the overall trends of molecular absorption strengths in the low-resolution IRS data. However, as we discuss in Section~\ref{sec:silicate_variability}, we find a weak correspondence between the presence of silicate absorption and known photometric variability.

\subsection{Dust: Silicates}
\label{sec:silicate}

In substellar atmospheres the silicate compound that is expected to condense at highest temperatures is forsterite (Mg$_2$SiO$_4$), at $T\approx1700$~K \citep[][at $P=1$~bar]{Lodders2002}. The condensation temperature increases or decreases, respectively, by $\approx$100~K for $P=10$~bar or 0.1 bar. At $\approx$50--100~K cooler temperatures, the SiO present in (equilibrium) atmospheres and the forsterite react to form enstatite (MgSiO$_3$; \citealt{Lodders2002}). The 1600--1700~K condensation temperatures of forsterite and enstatite correspond to mid-L spectral types \citep[e.g.;][]{Filippazzo_etal2015}.

Silicate molecules are known to have prominent broad spectroscopic signatures in the 8--12~$\micron$ range \citep[e.g.,][]{Draine_Lee1984,Hanner_etal1994}. Previous analyses of \textit{Spitzer} IRS spectra have already noted the presence of a 9--11~$\micron$ plateau in several L dwarfs. \citet{Cushing_etal2006} found that a combination of enstatite and forsterite condensates with small particle sizes ($\lesssim$2~$\micron$ radii) is able to reproduce the 9--11~$\micron$ flattening in the IRS spectrum of the L4.5 dwarf 2224$-$0158 (Figure \ref{fig:spectra_SL_2}). They observed a similar plateau in the IRS spectrum of 0036+1821 (L3.5) and a similar but weaker feature in 1507$-$1627 (L5). \citet{Looper_etal2008b} also identified this feature in the IRS spectra of 1821+1414 and 2148+4003 (L4.5 and L6; Figure \ref{fig:spectra_SL_2}) and attributed it to silicate absorption. \citet{Helling_etal2006} posit that the 9--11~$\micron$ feature could arise from quartz (SiO$_2$) grains, and note that it is absent in equilibrium models. They suggest that the detection of this feature could provide evidence for non-equilibrium dust formation in brown dwarf atmospheres. Accordingly, \citet{Burningham_etal2021} reported that a combination of enstatite and quartz condensates with iron clouds at different pressure layers reproduces well the 9 to 11~$\micron$ flattening in the IRS spectrum of 2224$-$0158. 

In our full IRS sample, we observe that a broad absorption feature is present in some L dwarfs between 8--11~$\micron$. We refer to this feature as ``silicate'' absorption, which we take to encompass the signatures of enstatite, forsterite, or quartz grains. Our ensemble analysis of all low-resolution \textit{Spitzer} IRS spectra shows that silicate absorption sets in as early as the L2 spectral type and is discernible through the end of the L dwarf sequence. 

\subsubsection{Silicate Absorption Strengths: Emergence and Sedimentation of Silicate Condensates in L Dwarfs}
\label{sec:silicate_absorption_in_Ldwarfs}

\begin{figure}
	\centering
	\includegraphics[width=1.0\linewidth]{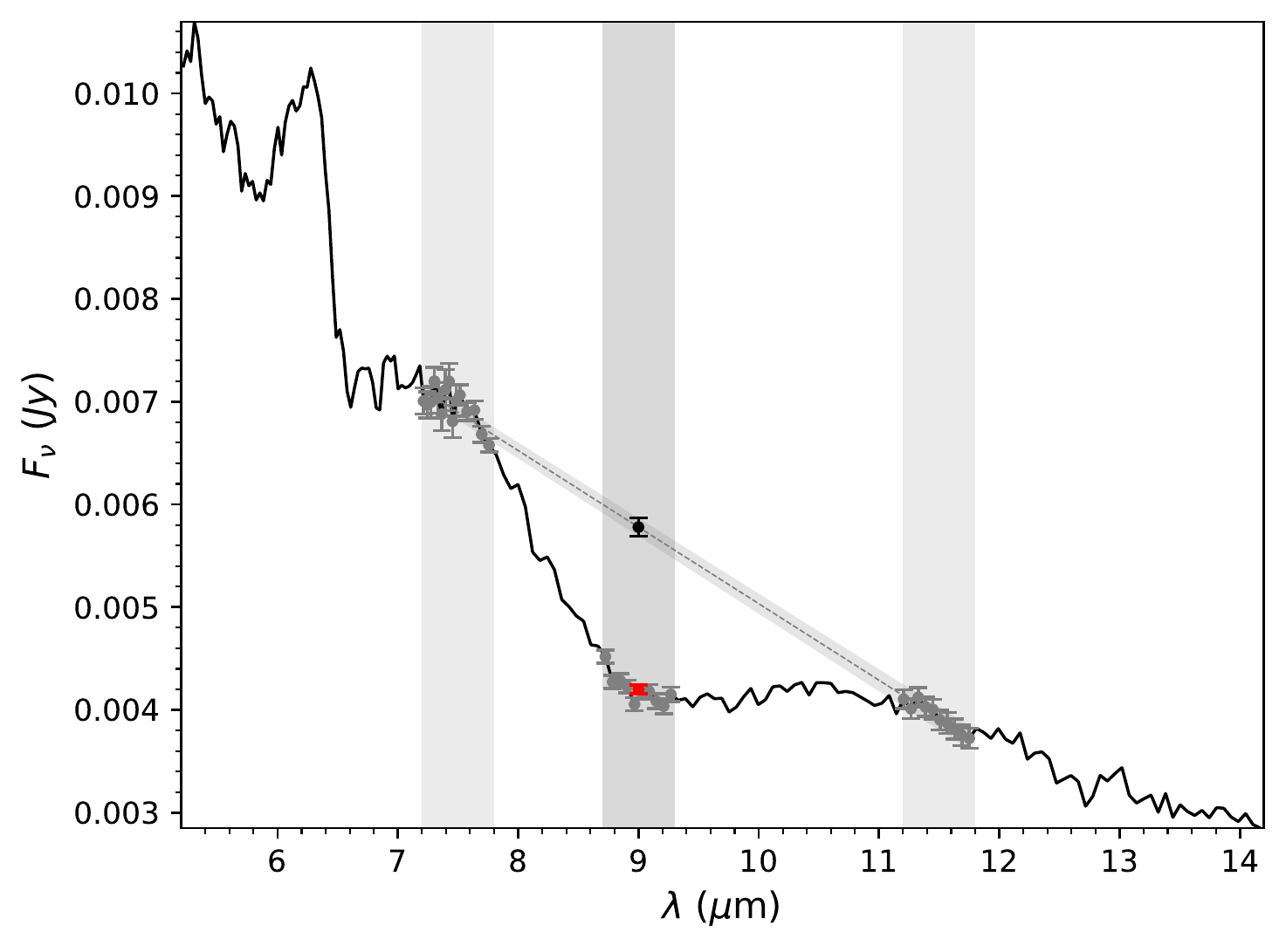}
	\caption{Example of how the silicate index was measured considering the IRS spectrum of the L4.5 dwarf 1821+1414. The shaded regions show the windows used to measure the silicate index (dark grey for the absorption and light grey for fitting the continuum). The black point indicates the interpolated continuum flux at 9.0~$\micron$ obtained by considering the best linear fit (black dashed line with uncertainties shown by the diagonally shaded region) to the fluxes at 7.5 and 11.5~$\micron$ in a window of 0.6~$\micron$ (grey points). The red point corresponds to the average flux at 9.0~$\micron$ in a same-width window (grey points). The silicate index is defined as the ratio of the interpolated continuum flux to the absorption flux.}
	\label{fig:silicate_example}
\end{figure}

We assessed the absorption strength over 8--11~$\micron$ in all IRS spectra of M5--T8 dwarfs. We defined the silicate index as the ratio of the interpolated continuum flux ($C_{9.0}$) at 9~$\micron$ to the average flux ($F_{9.0}$) in a window of 0.6~$\micron$ centred at 9~$\micron$. The interpolated continuum flux was obtained considering the best linear fit to the fluxes at 7.5 and 11.5~$\micron$ in a 0.6~$\micron$ window, as shown in Figure \ref{fig:silicate_example}. Thus, the index is given by 
\begin{ceqn}
\begin{align}
	{\rm Silicate\ Index} = \frac{C_{9.0}}{F_{9.0}}
	\label{eq:NH3}
\end{align}
\end{ceqn}

As for the water index in Equation~(\ref{eq:H2O}), we did not compute the silicate index for the five latest T dwarfs (Table \ref{tab:param}) because they do not have full spectral coverage in the region defined to measure the continuum of the feature. The silicate index uncertainties were obtained from the uncertainties in the mean flux density of the absorption and in the continuum fit. We list the silicate index measurements in Table~\ref{tab:indices}, and we present them as a function of spectral type in the bottom panel of Figure \ref{fig:indices}.

The trend in silicate index measurements over M5--T8 dwarfs, traced by the dashed curve in the bottom panel of Figure \ref{fig:indices}, shows a clear upward deviation in the L dwarfs. The above-trend silicate index values of L dwarfs indicate prevalent silicate absorption. The feature is discernible in the spectra of objects as early as spectral type L2 and as late as L8, and peaks in strength in L4--L6 dwarfs. At the same time, at all L subtypes---even at mid-L---there are objects that do not show any discernible silicate absorption. The diversity in silicate absorption strengths within a narrow spectral type range points to active sedimentation of different-particle size silicate condensates to deeper atmospheric layers and higher pressures, where they fall below the $\lesssim$1~bar 8--11~$\micron$ photosphere \citep{Cushing_etal2006,Karalidi_etal2021}.

The temperatures corresponding to the spectral types for the onset (L2) and disappearance (L8) of the silicate absorption in the IRS spectra are 1960~K and 1340~K, respectively, while the spectral type range where the feature is on average the strongest (L4--L6) corresponds to temperatures between 1690~K and 1480~K \citep{Filippazzo_etal2015}. The polynomial relationship between effective temperature and spectral type in \citet[][Table~10]{Filippazzo_etal2015} has an r.m.s.\ scatter of 113~K. A $\approx$2000~K onset temperature for silicate absorption is consistent with the recent finding that silicate condensation must begin between equilibrium temperatures 1950~K $<T_{\rm eq}<$ 2450~K in exoplanetary atmospheres \citep{Lothringer_etal2022}. Specifically, \citet{Lothringer_etal2022} detect the gas-phase SiO constituent of silicate molecules (MgSiO$_3$, Mg$_2$SiO$_4$) in the UV transmission spectra of $T_{\rm eq}=2350$~K and $T_{\rm eq}=2450$~K transiting planets, but do not see it in the UV transmission spectrum of a cooler $T_{\rm eq}=1950$~K transiting planet. Our result confirms that at this cooler temperature SiO would already be bound in silicates.

It is possible that silicates may start condensing at slightly earlier spectral types and warmer temperatures. There are three M9.5--L1 (2320~K $\ge T_{\rm eff}\ge$ 2100~K) dwarfs (1013$-$1356, 0911+7401, and 1022+5825) with a high silicate index ($>$1.2). However, they also have large ($\ge$20 per cent) silicate index uncertainties due to the low S/N ($<$5 at 12~$\micron$) of their IRS spectra, such that the detection of the feature is not significant. None of the $\leq$M9 ($T_{\rm eff}\geq2400$~K) dwarfs or $\ge$T0 ($T_{\rm eff}\leq1250$~K) dwarfs show any silicate absorption at the S/N$\lesssim$100 and $R\approx90$ resolution of most archival IRS spectra.

\subsubsection{Positive Correlation between Silicate Absorption and Near-infrared Colour Excess}
\label{sec:silicate_NIRcolours}
It is established that certain L dwarfs with prominent silicate absorption in their IRS spectra (1821+1414, 2148+4003, and 2224$-$0158) have redder colours than expected for their spectral types \citep{Cushing_etal2006,Cushing_etal2008,Looper_etal2008b}. \citet{Looper_etal2008b} suggested that the redness of 1821+1414 and 2148+4003 is due to unusually dusty atmospheres caused by high metallicity or low surface gravity. If the colour peculiarities of L dwarfs at a given subtype are caused by condensate silicate clouds, a correlation between colours and the absorption strength of the Si-O feature could be expected \citep{Burgasser_etal2008b,Cushing_etal2008,Looper_etal2008b}. We investigate this correlation with our IRS spectra in the following.

We selected the dwarfs with the largest ($>$1.2) silicate index values---above the dotted line in the bottom panel of Figure~\ref{fig:indices}---and with fractional errors $<$20 per cent. We visually verified in Figures~\ref{fig:spectra_SL_1} and \ref{fig:spectra_SL_2} that the selected dwarfs indeed show a flux deficit between 8 and 11~$\micron$. The resulting sample of 21 dwarfs spans the L3--L7 spectral subtypes, and includes the four objects previously reported to exhibit strong silicate features (0036+1821 and 2224$-$0158 in \citealt{Cushing_etal2006}, and 1821+1414 and 2148+4003 in \citealt{Looper_etal2008b}). Two other L dwarfs with previously reported weak silicate absorption, 0255$-$4700 \citep[L8;][]{Roellig_etal2004} and 1507$-$1627 \citep[L5;][]{Cushing_etal2006,Looper_etal2008b}, were not selected because their silicate index values are below our threshold. 

\begin{figure*}
	\centering
	\includegraphics[width=0.95\linewidth]{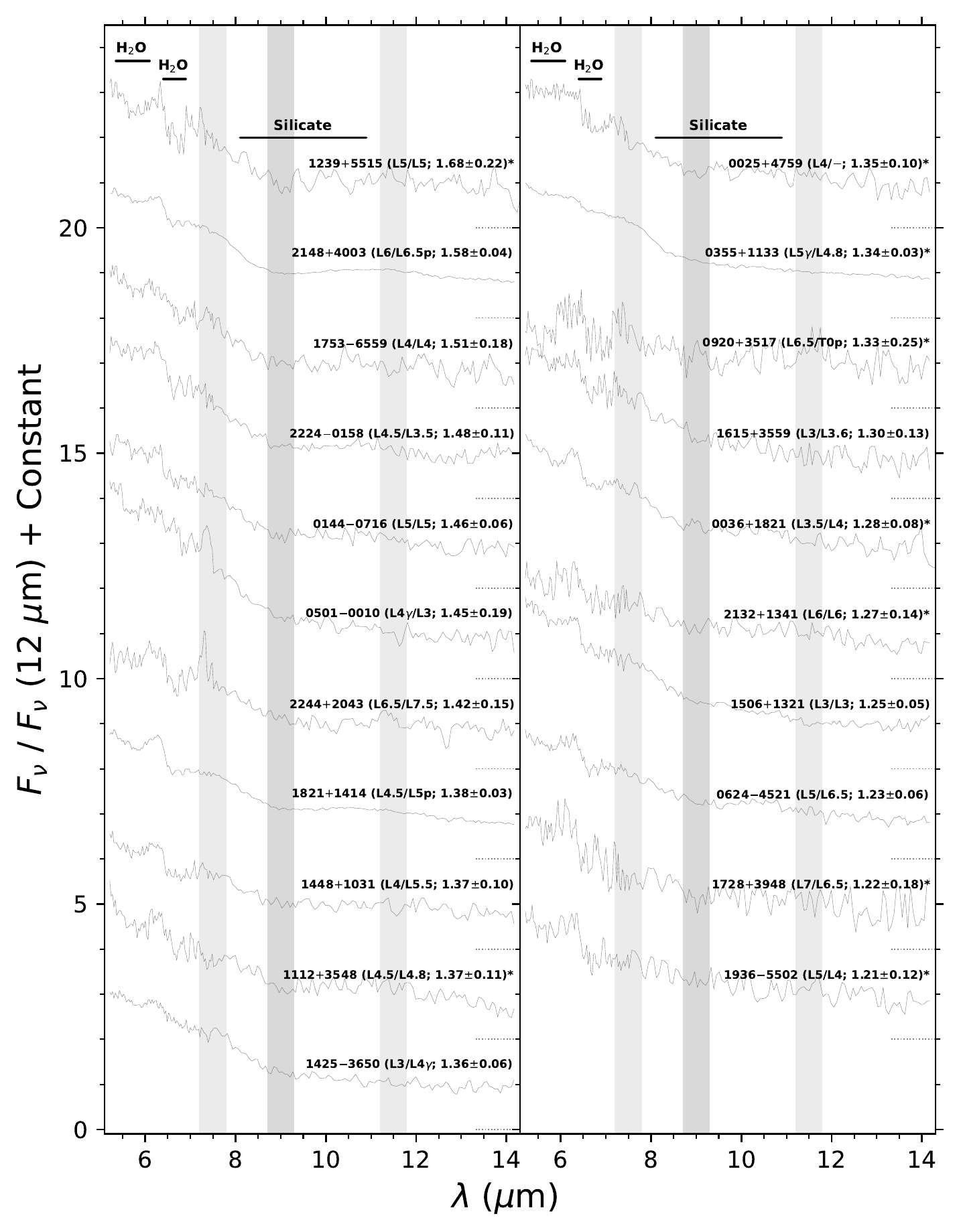}
    	\caption{IRS spectra of the dwarfs with the largest silicate index values ($\ge$1.2 and fractional errors <20 per cent). The selected dwarfs have (adopted optical) spectral types between L3 and L7. Optical and near-infrared spectral types (when available) are indicated for each object, as is the silicate index value and its uncertainty. The spectra were normalised to unity using the median flux at 12~$\micron$ in a 0.6~$\micron$ window and offset by constants (dotted lines). The shaded regions show the windows used to measure the silicate index (dark grey for the absorption and light grey for fitting the continuum). The spectra are ordered by decreasing silicate index from the top left to the bottom right. Unresolved binaries are marked with asterisks (Table \ref{tab:param}).}
	\label{fig:spectra_silicate}
\end{figure*}

Figure \ref{fig:spectra_silicate} shows the IRS spectra of the 21 L3--L7 dwarfs selected to have strong silicate absorption. The optical and near-infrared spectral types (when available) of these dwarfs are in agreement within one subtype, except for 0920+3517. This object has an adopted (Section \ref{sec:data_search}) optical spectral type L6.5 \citep{Kirkpatrick_etal2000}, but its infrared classification is T0 \citep{Burgasser_etal2006b}. It is in fact a known close binary with the individual components having near-infrared spectral types of L5.5 and L9 \citep{Dupuy-Liu2012}. As the binary is not resolved in the \textit{Spitzer} IRS beam, we are likely seeing the brighter L5.5 dwarf dominating the mid-infrared continuum.

In Figure \ref{fig:NIR_colors} we compare the 2MASS near-infrared colours of the L3--L7 dwarfs with the strongest silicate absorption (marked with red data points) to the median 2MASS colours of all dwarfs in the UltracoolSheet catalogue (marked with grey data points), as a function of spectral type. We observe that the dwarfs with the strongest silicate absorption are preferentially redder in all near-infrared colour combinations, even if they lie within the 68 per cent central confidence interval (grey band) for most cases. 

We further investigate this behaviour by comparing the silicate index values of 37 of the 39 L3--L7 dwarfs with IRS spectra with their 2MASS near-infrared colour excesses. We define the colour excess of an object as the difference between its observed colour and the median colour for its spectral type (Figure~\ref{fig:NIR_colors}). Working with colour excesses rather than just colours removes the dependence on spectral type and isolates the potential role of reddening by silicate dust. The two mid-L objects excluded from this analysis are the subdwarfs 0532+8246 (sdL7) and 1626+3925 (sdL4) because they have very blue 2MASS colours (Figure \ref{fig:NIR_colors}). (As the only L subdwarfs in our sample, these two objects are expected to have condensate-free atmospheres, so we are justified in excluding them from our analysis of silicate absorption strengths.) Figure \ref{fig:silicate_vs_NIRcolor} shows this comparison, and reveals a positive correlation between silicate absorption strength and 2MASS colour excess. That is, L3--L7 dwarfs that are redder than average exhibit a stronger silicate absorption, while bluer dwarfs have a weak or absent absorption. The Pearson correlation coefficients $r$ are between 0.5 and 0.7 with $p$-values of $<$0.2 per cent: indicative of a moderate to strong correlation \citep{Cohen1988}. We verified that similar but less significant ($0.4\le r \le0.5$) correlations hold for all L0--L8 dwarfs with IRS spectra (excluding the two L subdwarfs). 

This result is consistent with the scenario where the near-infrared colour scatter (including peculiar colours) at a given L subtype is caused by differences in dust cloud thickness \citep{Knapp_etal2004,Burgasser_etal2008b,Cushing_etal2008,Looper_etal2008b}. Specifically, L dwarfs in which the silicate condensate column extends into the $<$1~bar upper atmosphere, where it is detectable with \textit{Spitzer} IRS, tend to on average have redder near-infrared colours.

\subsubsection{Silicate Absorption vs.\ Photometric Variability: Variable mid-L Dwarfs Are Preferentially Silicate-rich Objects}
\label{sec:silicate_variability}
Photospheric clouds are considered the leading mechanism for photometric variability in L and T dwarfs \citep[e.g.,][and references therein]{Burgasser_etal2002,Marley_etal2010,Radigan_etal2014,Miles-Paez_etal2017c,Biller2017}. \citet{Apai_etal2013} further note that photometric variability in L and T dwarfs is associated with near-infrared colour excursions along a dust extinction track: consistent with varying cloud thickness if the clouds are composed of dust. It could thus be anticipated that if dust clouds are responsible for the observed photometric variability in L dwarfs, then the presence of silicate absorption could be related to observed variability.

\begin{figure}
	\centering
	\includegraphics[width=1.0\linewidth]{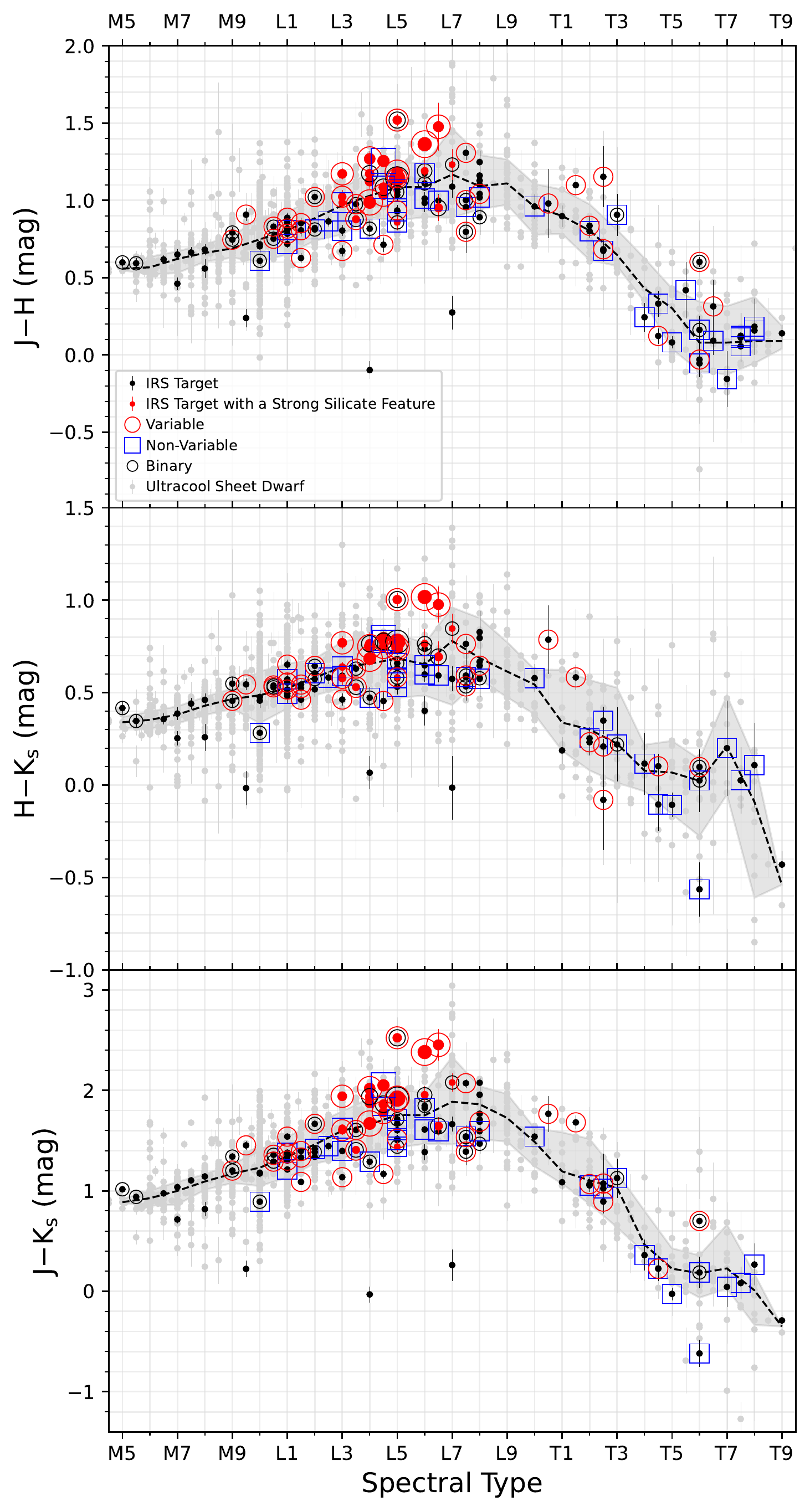}
	\caption{Near-infrared 2MASS colours as a function of spectral type for all $>$2000 M5--T9 dwarfs in the UltracoolSheet catalogue (grey points). All field M5--T9 dwarfs with IRS spectra and 2MASS photometry are shown in with black points, while those with the largest silicate index values (see Fig.~\ref{fig:indices}, bottom panel) are shown with red points (with the symbol size proportional to the index value). The red circles, blue squares, and black circles indicate variable, non-variable, and binary sources (Table \ref{tab:param}), respectively. The dashed curve and the shaded area in each panel represent the median index and 68 per cent central confidence interval for all dwarfs in the UltracoolSheet catalogue, calculated in bins of one spectral subtype.}
	\label{fig:NIR_colors}
\end{figure}

\begin{figure*}
	\centering
	\includegraphics[width=1.0\linewidth]{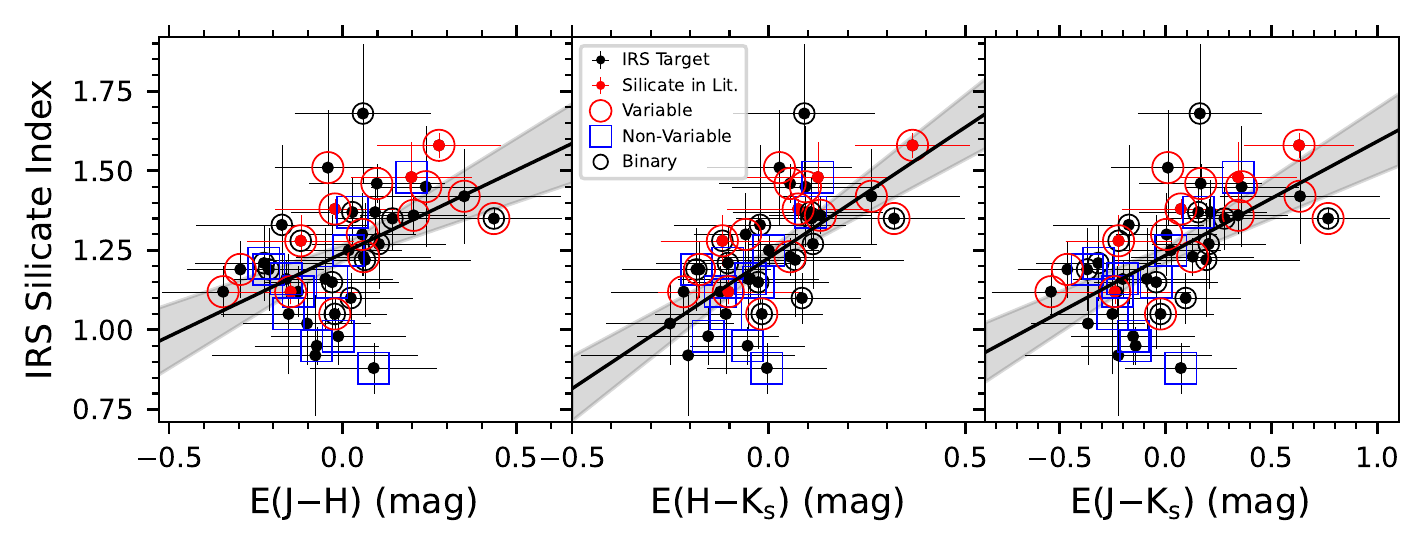}
	\caption{Silicate index as a function of 2MASS colour excess (Section~\ref{sec:silicate_NIRcolours}) for 37 of the 39 L3--L7 dwarfs with IRS spectra. Among these, the red points indicate the five mid-L objects with previously-reported silicate absorption: 0036+1821 (L3.5), 1507$-$1627 (L5), and 2224$-$0158 (L4.5) in \citet{Cushing_etal2006}, and 1821+1414 (L4.5) and 2148+4003 (L6) in \citet{Looper_etal2008b}. The red circles, blue squares, and black circles indicate variable, non-variable, and binary sources (Table \ref{tab:param}), respectively. The best linear fit to the data points in each panel is indicated by the black line. The shaded region represents the 1$\sigma$ uncertainties of the fit in both slope and intercept. We excluded the subdwarfs 0532+8246 (sdL7) and 1626+3925 (sdL4) that have very blue 2MASS colours (e.g., $J-K_s<0.5$~mag; Figure \ref{fig:NIR_colors}).}
	\label{fig:silicate_vs_NIRcolor}
\end{figure*}

Thirty-nine of the field M5--T9 dwarfs in the \textit{Spitzer} IRS sample have been reported to be photometrically variable from previous $<$5~$\micron$ observations (considering the compilation of L and T variables in \citealt{Tannock_etal2021} and complementing with additional references, as listed in Table~\ref{tab:param}). We can test whether L dwarfs displaying mid-infrared silicate absorption are also more likely to be known as variables. We focus specifically on the 37 L3--L7 dwarfs, among which we have the 21 objects with the strongest silicate absorption (Section~\ref{sec:silicate_NIRcolours}; excluding the two L subdwarfs). Out of these 37 L3--L7 dwarfs, 27 have been monitored for variability, with 15 detected variables and 12 non-variables. For the remaining 10 we could not find literature references to variability monitoring.  All variables are marked with red circles in Figures~\ref{fig:indices}, \ref{fig:NIR_colors}, and \ref{fig:silicate_vs_NIRcolor}, and known non-variables are marked with blue squares.

\begin{figure*}
	\centering
	$\vcenter{\hbox{\includegraphics[width=0.30\linewidth]{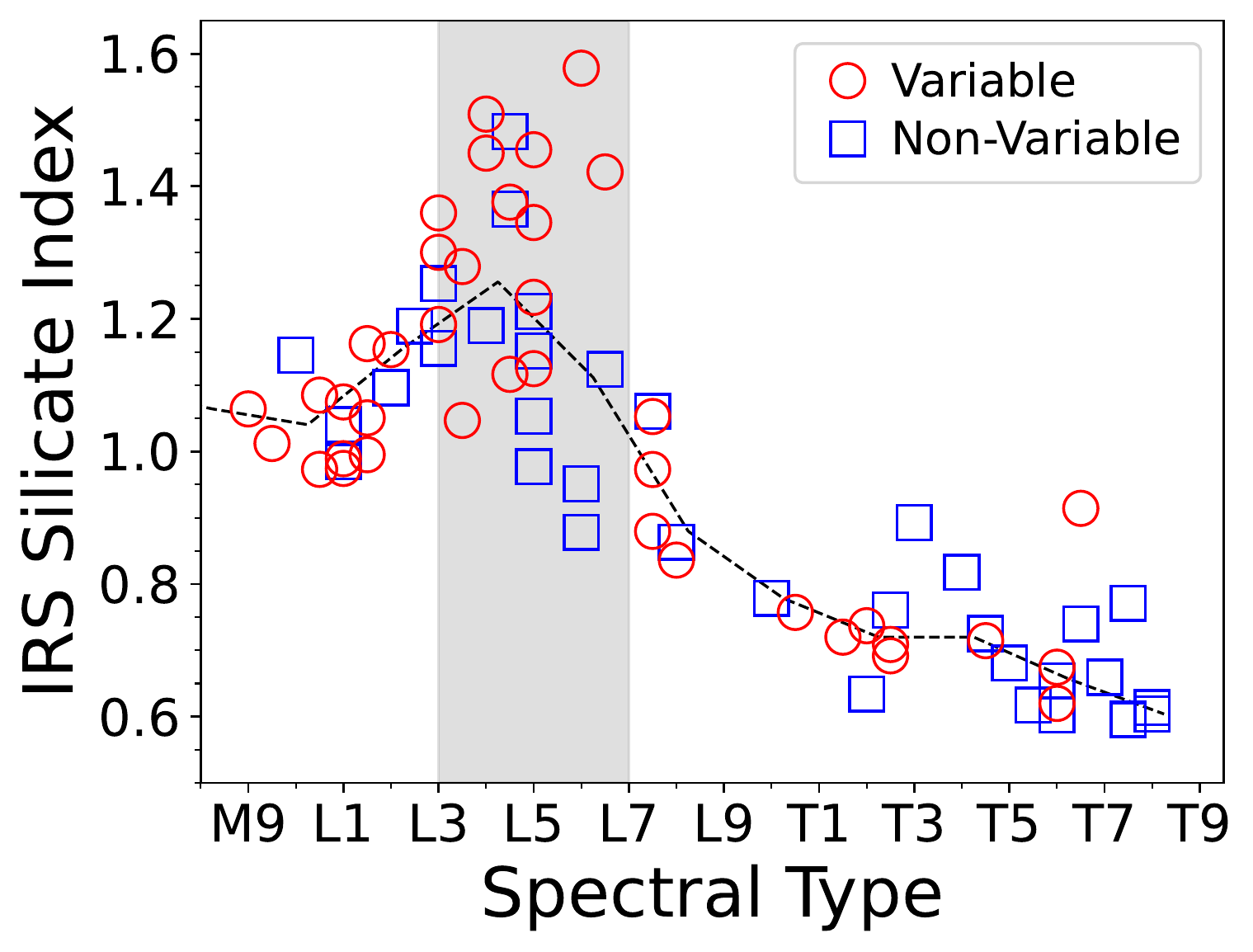}}}$
	$\vcenter{\hbox{\includegraphics[width=0.69\linewidth]{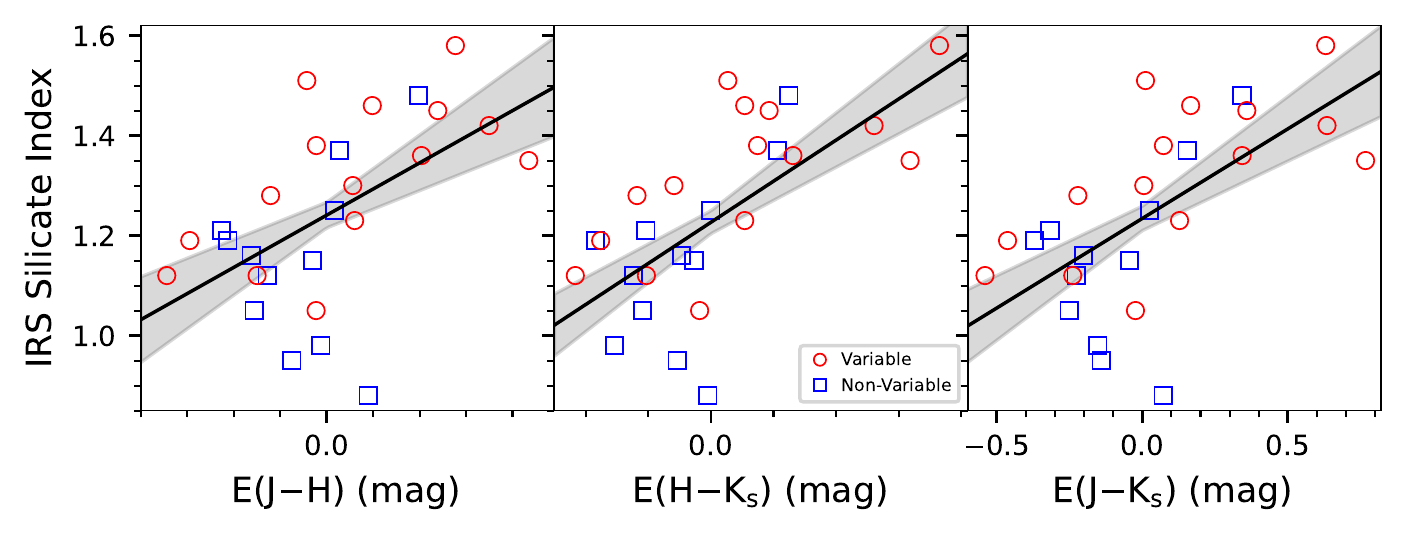}}}$
    	\caption{Silicate index as a function of spectral type (left panel) or near-infrared 2MASS colour excess (right three panels) for variable (red circles) and non-variable (blue squares) ultracool dwarfs with IRS spectra (Table~\ref{tab:param}). The shaded region in the left panel indicate the L3--L7 spectral type range where we focus the analysis on variability. The dashed curve in the left panel, and the solid line and shaded region in the right three panels, are the same as in Figures~\ref{fig:indices} (bottom panel) and \ref{fig:silicate_vs_NIRcolor}, respectively. The right three panels focus on L3--L7 dwarfs, for which the presented assessment of variability and non-variability from the published literature is complete. We have excluded the two L subdwarfs (1626+3925 and 0532+8246), as explained in Section~\ref{sec:silicate_NIRcolours}.}
	\label{fig:silicate_vs_var}
\end{figure*}

For clarity, we have extracted the variability information in a separate Figure~\ref{fig:silicate_vs_var}.  The left panel of Figure~\ref{fig:silicate_vs_var} is a simplified version of the bottom panel of Figure~\ref{fig:indices}, that contains only the variability indications. Similarly, the right three panels of Figure~\ref{fig:silicate_vs_var} are simplified versions of the three panels in Figure~\ref{fig:silicate_vs_NIRcolor}.

Of the 15 variable L3--L7 dwarfs, 11 have above-average silicate absorption for their spectral types (dark dashed curve in the left panel of Figure~\ref{fig:silicate_vs_var}) and four are below average. If silicate absorption was independent of variability, the binomial probability of having 11 or more variables with above-average silicate strength is 5.9 per cent. Conversely, of the 12 known non-variable L3--L7 dwarfs, only four have above-average silicate strengths, and eight have weaker silicate. Again, the binomial probability of having at most four variables with above-average silicate occurring at random is 19.4 per cent. Neither of these probabilities are very low individually, but the joint occurrence of the phenomena is even less probable. These two distributions are not independent, because they jointly contribute to the calculation of the average silicate absorption strength. To properly assess the joint probability of randomly having at least 11 silicate-rich objects among the 15 variables, and at most 4 silicate-rich objects among the 12 non-variables, we performed a Monte Carlo simulation with $10^6$ realisations of a sample of 37 L3--L7 dwarfs (of which 15 are variables, 12 are non-variables, and 10 are with an unknown variability status). In only 1.2 per cent of cases did a similar or more extreme distribution occur at random. Hence, our analysis shows that silicate absorption and photometric variability are co-dependent with 98.8 per cent significance.

The comparison between silicate absorption strength and near-infrared colour excess is less indicative of a correlation. In the excess analysis for each near-infrared colour, ten of the fifteen known variable L3--L7 dwarfs have above-average silicate strengths, as seen from the right three panels of Figure~\ref{fig:silicate_vs_var}. Among the 12 non-variables, seven are below average in all colour excess combinations. The former occurrence is significant at the 85 per cent level, whereas the latter is consistent with being random.

Overall, we find a probable (98.8 per cent) indication that variable L3--L7 dwarfs are $\approx$2 times more likely to exhibit above-average silicate absorption than non-variable L3--L7 dwarfs (11/15 for variables vs.\ 4/12 for non-variables). If dependence on spectral type in the L3--L7 range is ignored and only near-infrared colour excess is considered, the significance of the correlation is weaker (85 per cent). The probable link between variability and silicate absorption strength affirms earlier suggestions \citep{Burgasser_etal2002, Marley_etal2010} that the clouds responsible for photometric variability in L dwarfs are composed of silicates. 

\section{Summary and Conclusions}
\label{sec:summary_conclusions}
We have reprocessed and presented $R\approx90$ mid-infrared spectra of all 113 field M5--T9 dwarfs observed and detected with the \textit{Spitzer} IRS. These include 12 M5--M9 dwarfs, 69 L dwarfs, and 32 T dwarfs (Table~\ref{tab:param}). Spectra for 68 of the 113 (i.e., 60 per cent) detected field M5--T9 dwarfs are presented here for the first time, although 32 of the 68 spectra are mentioned in previous references without being shown or available (Table~\ref{tab:log}). Most of the spectra cover the 5.2--14.2~$\micron$ wavelength range, except for the spectra of the five latest T dwarfs, which cover 7.5--14.2~$\micron$  (Figures~\ref{fig:spectra_SL_1}, \ref{fig:spectra_SL_2}, and \ref{fig:spectra_SL_3}). Eighteen of the spectra have coverage that extends up to 38~$\micron$ (Figure~\ref{fig:spectra_SL_LL}).

We have analysed the strengths of the main mid-infrared molecular and condensate absorbers, namely H$_2$O at 6.25~$\micron$, CH$_4$ at 7.65~$\micron$, NH$_3$ at 10.5~$\micron$, and silicates at 8--11~$\micron$. Water absorption is present in all spectra and strengthens with increasing spectral type. Methane and ammonia features appear consistently at L8 and T2.5, respectively, although ammonia may be discernible as early as T1.5. These findings are consistent with the analysis of select M, L, and T IRS spectra from \citet{Cushing_etal2006}, and improve the robustness of these early results.

Our uniform analysis of all field M5--T9 IRS spectra traces for the first time the emergence and sedimentation of silicate condensates in L dwarfs. To this end we design an index to measure the strength of the 8--11~$\micron$ flux deficit previously attributed to silicate absorption in select L dwarfs. Our analysis, bound by the S/N ratio and resolution limits of the archival \textit{Spitzer} IRS spectra, shows that silicate condensates first attain measurable optical depth in the visible atmosphere of ultracool dwarfs at spectral type L2 ($T_{\rm eff}=1960$~K), are thickest between L4--L6 (1690~K $\geq T_{\rm eff} \geq$ 1480~K), and sediment below the 8--11~$\micron$ photosphere past spectral type L8 ($T_{\rm eff}=1340$~K; bottom panel of Figure~\ref{fig:indices}). However, at any L subtype silicate absorption can also be absent from the IRS spectra, even for mid-L dwarfs. 

Objects with the strongest silicate absorption are preferentially redder in the near-infrared compared to other dwarfs with similar spectral types (Figure~\ref{fig:NIR_colors}). Specifically, there is a significant correlation between the silicate absorption strength and the deviation from median near-infrared 2MASS colours (colour excess) of L3--L7 dwarfs (Figure~\ref{fig:silicate_vs_NIRcolor}), as suggested in previous studies \citep{Burgasser_etal2008b,Cushing_etal2008,Looper_etal2008b}. A similar positive correlation holds, but with less significance, for all L dwarfs with IRS spectra. This result is consistent with the scenario in which variations of condensate cloud thickness produce the observed colour scatter at a fixed subtype in L dwarfs \citep{Knapp_etal2004,Burgasser_etal2008b,Cushing_etal2008,Looper_etal2008b}. 

Variable mid-L dwarfs exhibit on average stronger silicate absorption than non-variables with similar spectral types or 2MASS colour excesses (Figure~\ref{fig:silicate_vs_var}). Conversely, L3--L7 dwarfs with above-average silicate absorption are approximately twice more likely to be photometrically variable.

The consistent detection of silicate absorption in the IRS L dwarf sample, its progression in strength, its correlation with near-infrared colour excess, and its $\approx$99 per cent probable link to photometric variability, solidify the evidence that: (1) silicates condense to form clouds in L dwarf atmospheres, (2) these silicate clouds govern L-dwarf variability, and (3) they are responsible for the observed colour scatter in L dwarfs. The photometric variability of early T dwarfs is also likely caused by silicate clouds, but the lack of detectable silicate absorption in early-T IRS spectra indicates that the silicates have sedimented to higher pressure levels than the $\lesssim$1~bar probed in the 8--11~$\micron$ wavelength range.

There are many other possible applications of the \textit{Spitzer} IRS spectroscopic sample presented here \citep[e.g.;][]{Cushing_etal2006,Burningham_etal2021,Suarez_etal2021a}. We expect that this sample will be used in future studies to improve our understanding of the atmospheric physics and chemistry of ultracool dwarfs, and will serve as a basis for planning and interpreting ultracool dwarf and exoplanet observations with the \textit{James Webb Space Telescope}.

\bigskip

\section*{Acknowledgements}
We thank Theodora Karalidi for providing an extended (to 25~$\mu$m) version of the plot of probed atmospheric pressure vs.\ wavelength published in \citet{Karalidi_etal2021}. Support for this work was provided by the Canadian Space Agency under the 2019 announcement of opportunity for the Flights and Fieldwork for the Advancement of Science and Technology (FAST-AO 2019) programme (reference no.\ 19FAWESB40). The spectra presented here are obtained entirely with the \textit{Spitzer} Space Telescope, which is operated by the Jet Propulsion Laboratory, California Institute of Technology under a contract with NASA. This work has benefitted from The UltracoolSheet, maintained by Will Best, Trent Dupuy, Michael Liu, Rob Siverd, and Zhoujian Zhang, and developed from compilations by \citet{Dupuy-Liu2012}, \citep{Dupuy-Kraus2013}, \citet{Liu_etal2016}, \citet{Best_etal2018}, and \citet{Best_etal2021}.

\section*{Data availability}
The data underlying this article are available in the article and in its online supplementary material and on Genaro Su\'arez' homepage\footnote{\url{https://gsuarezcastro.wixsite.com/gsuarez/research}}.

\defcitealias{Albert_etal2011}{Albe11}
\defcitealias{Allers-Liu2013}{Alle13}
\defcitealias{Aller_etal2016}{Alle16}
\defcitealias{Artigau_etal2006}{Arti06}
\defcitealias{Artigau_etal2009}{Arti09}
\defcitealias{Bailer-Jones-Mundt2001}{Bail01}
\defcitealias{Bailer-Jones-Lamm2003}{Bail03}
\defcitealias{Bailer-Jones2004}{Bail04}
\defcitealias{BardalezGagliuffi_etal2014}{Bard14}
\defcitealias{Bartlett_etal2017}{Bart17}
\defcitealias{Bernat_etal2010}{Bern10}
\defcitealias{Bessell1991}{Bess91}
\defcitealias{Boeshaar1976}{Boes76}
\defcitealias{Bouy_etal2003}{Bouy03}
\defcitealias{Buenzli_etal2014}{Buen14}
\defcitealias{Burgasser_etal1999}{Burg99}
\defcitealias{Burgasser_etal2000}{Burg00a}
\defcitealias{Burgasser_etal2000b}{Burg00b}
\defcitealias{Burgasser_etal2002}{Burg02}
\defcitealias{Burgasser_etal2003c}{Burg03a}
\defcitealias{Burgasser_etal2003d}{Burg03b}
\defcitealias{Burgasser_etal2003b}{Burg03c}
\defcitealias{Burgasser_etal2003}{Burg03d}
\defcitealias{Burgasser2004c}{Burg04a}
\defcitealias{Burgasser_etal2004}{Burg04b}
\defcitealias{Burgasser_etal2005b}{Burg05a}
\defcitealias{Burgasser_etal2005}{Burg05b}
\defcitealias{Burgasser_etal2006}{Burg06a}
\defcitealias{Burgasser_etal2006b}{Burg06b}
\defcitealias{Burgasser_etal2006c}{Burg06c}
\defcitealias{Burgasser_etal2007}{Burg07}
\defcitealias{Burgasser_etal2008b}{Burg08a}
\defcitealias{Burgasser_etal2008c}{Burg08b}
\defcitealias{Burgasser_etal2010b}{Burg10}
\defcitealias{Burningham_etal2008}{Burn08}
\defcitealias{Burningham_etal2009}{Burn09}
\defcitealias{Chiu_etal2006}{Chiu06}
\defcitealias{Clarke_etal2002}{Clar02}
\defcitealias{Clarke_etal2002b}{Clar02b}
\defcitealias{Cruz_etal2003}{Cruz03}
\defcitealias{Cruz_etal2007}{Cruz07}
\defcitealias{Cruz_etal2009}{Cruz09}
\defcitealias{Dahn_etal2002}{Dahn02}
\defcitealias{Dahn_etal2008}{Dahn08}
\defcitealias{Delorme_etal2008}{Delo08}
\defcitealias{Dieterich_etal2014}{Diet14}
\defcitealias{Dupuy-Liu2012}{Dupu12}
\defcitealias{Dupuy_etal2015}{Dupu15}
\defcitealias{EROSCollaboration1999}{EROS99}
\defcitealias{Fan_etal2000}{Fan00}
\defcitealias{Folkes_etal2007}{Folk07}
\defcitealias{Gagne_etal2015}{Gagn15}
\defcitealias{Geballe_etal2002}{Geba02}
\defcitealias{Gelino_etal2002}{Geli02}
\defcitealias{Girardin_etal2013}{Gira13}
\defcitealias{Gizis1997}{Gizi97}
\defcitealias{Gizis_etal2000}{Gizi00}
\defcitealias{Gizis2002}{Gizi02}
\defcitealias{Gizis_etal2003}{Gizi03}
\defcitealias{Golimowski_etal2004b}{Goli04}
\defcitealias{Harding_etal2013}{Hard13}
\defcitealias{Hawley_etal2002}{Hawl02}
\defcitealias{Henry-Kirkpatrick1990}{Henr90}
\defcitealias{Irwin_etal1991}{Irwi91}
\defcitealias{Kao_etal2018}{Kao18}
\defcitealias{Kendall_etal2004}{Kend04}
\defcitealias{Kirkpatrick_etal1991}{Kirk91}
\defcitealias{Kirkpatrick_etal1995}{Kirk95}
\defcitealias{Kirkpatrick_etal1999}{Kirk99}
\defcitealias{Kirkpatrick_etal2000}{Kirk00}
\defcitealias{Kirkpatrick_etal2001}{Kirk01}
\defcitealias{Kirkpatrick_etal2008}{Kirk08}
\defcitealias{Kirkpatrick_etal2010}{Kirk10}
\defcitealias{Knapp_etal2004}{Knap04}
\defcitealias{Koen2003}{Koen03}
\defcitealias{Koen_etal2004}{Koen04}
\defcitealias{Koen2013}{Koen13}
\defcitealias{Leggett_etal1996}{Legg96}
\defcitealias{Leggett_etal2000}{Legg00}
\defcitealias{Leinert_etal1986}{Lein86}
\defcitealias{Lepine_etal2003}{Lepi03}
\defcitealias{Liebert_etal2003}{Lieb03}
\defcitealias{Liu-Leggett2005}{Liu05}
\defcitealias{Lodieu_etal2005}{Lodi05}
\defcitealias{Looper_etal2007}{Loop07}
\defcitealias{Looper_etal2008b}{Loop08}
\defcitealias{Luhman_etal2007}{Luhm07a}
\defcitealias{Luyten1949}{Luyt49}
\defcitealias{Miles-Paez_etal2017b}{Mile17}
\defcitealias{Marocco_etal2013}{Maro13}
\defcitealias{Marocco_etal2015}{Maro15}
\defcitealias{Martin_etal1999}{Mart99}
\defcitealias{Martin_etal2001}{Mart01}
\defcitealias{McCaughrean_etal2004}{McCa04}
\defcitealias{McElwain-Burgasser2006}{McEl06}
\defcitealias{Metchev_etal2015}{Metc15}
\defcitealias{Morales-Calderon_etal2006}{Mora06}
\defcitealias{Pineda_etal2016}{Pine16}
\defcitealias{Pokorny_etal2004}{Poko04}
\defcitealias{Pope_etal2013}{Pope13}
\defcitealias{Radigan_etal2012}{Radi12}
\defcitealias{Radigan_etal2014}{Radi14}
\defcitealias{Reid_etal2000}{Reid00}
\defcitealias{Reid_etal2001}{Reid01}
\defcitealias{Reid_etal2006}{Reid06}
\defcitealias{Reid_etal2008}{Reid08}
\defcitealias{Ruiz_etal1997}{Ruiz97}
\defcitealias{Schneider_etal2014}{Schn14}
\defcitealias{Scholz_etal2003}{Scho03}
\defcitealias{Scholz_etal2004}{Scho04}
\defcitealias{Siegler_etal2007}{Sieg07}
\defcitealias{Strauss_etal1999}{Stra99}
\defcitealias{Stumpf_etal2008}{Stum08}
\defcitealias{Thorstensen-Kirkpatrick2003}{Thor03}
\defcitealias{Tinney_etal2005}{Tinn05}
\defcitealias{vanBiesbroeck1961}{vanB61}
\defcitealias{Vos_etal2018}{Vos18}
\defcitealias{Vos_etal2020}{Vos20}
\defcitealias{Vos_etal2022}{Vos22}
\defcitealias{Warren_etal2007}{Warr07}
\defcitealias{Wilson_etal2001}{Wils01}
\defcitealias{Wilson_etal2003b}{Wils03}
\defcitealias{Wilson_etal2014}{Wils14}
\defcitealias{Zhou_etal2018}{Zhou18}

\clearpage
\onecolumn
\begin{landscape}
\begin{ThreePartTable}
\small
\begin{longtable}{@{\extracolsep{-2pt}}lcccccccccccc@{}}
	\caption{Select properties of the 113 field M5--T9 dwarfs with detectable signal in \textit{Spitzer} IRS spectra.} \label{tab:param} \\
	\toprule
	Designation              & Short Name      & Other Name                & R.A.$_{\textrm{J2000}}$       & Decl.$_{\textrm{J2000}}$      & Discovery Ref.                          & SpT$_{\textrm{opt}}$ & Ref.                                     & SpT$_{\textrm{IR}}$ & Ref.                                    & Binary                                & Variable                               & Non-Variable                                     \\
	                         & (hhmm$\pm$ddmm) &                           & ($^\circ$)                    & ($^\circ$)                    &                                         &                      &                                          &                     &                                         &                                       &                                        &                                                  \\
	\midrule                                                                                                                                                                                                                                                                                                                                                                                                         
	\endfirsthead                                                                                                                                                                                                                                                                                                                                                                                                    
                                                                                                                                                                                                                                                                                                                                                                                                                     
	\multicolumn{13}{c}
	{{\bfseries \tablename\ \thetable{}} -- (Continued)} \\                                                                                                                                                                                                                                                                                                                                                          
	\toprule                                                                                                                                                                                                                                                                                                                                                                                                         
	Designation              & Short Name      & Other Name                & R.A.$_{\textrm{J2000}}$       & Decl.$_{\textrm{J2000}}$      & Discovery Ref.                          & SpT$_{\textrm{opt}}$ & Ref.                                     & SpT$_{\textrm{IR}}$ & Ref.                                    & Binary                                & Variable                               & Non-Variable                                     \\
	                         & (hhmm$\pm$ddmm) &                           & ($^\circ$)                    & ($^\circ$)                    &                                         &                      &                                          &                     &                                         &                                       &                                        &                                                  \\
	\midrule                                                                                                                                                                                                                                                                                                                                                                                                         
	\endhead                                                                                                                                                                                                                                                                                                                                                                                                         

	2MASS J00001354+2554180    & 0000+2554   & \nodata                   &  0.0563   & 25.9055     & \citetalias{Knapp_etal2004}              & T5                   & \citetalias{Pineda_etal2016}             & T4.5                & \citetalias{Burgasser_etal2006b}        & \nodata                               & \nodata                                & \citetalias{Buenzli_etal2014}       \\
	2MASS J00043484$-$4044058  & 0004$-$4044 & GJ 1001BC, LHS 102BC      &  1.1453   & $-$40.7351  & \citetalias{EROSCollaboration1999}       & L5                   & \citetalias{Kirkpatrick_etal2001}        & L4.5                & \citetalias{Knapp_etal2004}             & \citetalias{Golimowski_etal2004b}     & \nodata                                & \citetalias{Dieterich_etal2014}     \\
	2MASS J00242463$-$0158201  & 0024$-$0158 & DY Psc, BRI 0021-0214     &  6.1026   & $-$1.9722   & \citetalias{Irwin_etal1991}              & M9.5                 & \citetalias{Kirkpatrick_etal1995}        & M9.5                & \citetalias{Geballe_etal2002}           & \nodata                               & \citetalias{Martin_etal2001}           & \nodata                             \\
\bottomrule
\end{longtable}
\begin{tablenotes}[para,flushleft]
	{\bf References.} Albe11: \citet{Albert_etal2011}, Alle13: \citet{Allers-Liu2013}, Alle16: \citet{Aller_etal2016}, Arti06: \citet{Artigau_etal2006}, Arti09: \citet{Artigau_etal2009}, Bail01: \citet{Bailer-Jones-Mundt2001}, Bail03: \citet{Bailer-Jones-Lamm2003}, Bail04: \citet{Bailer-Jones2004}, Bard14: \citet{BardalezGagliuffi_etal2014}, Bart17: \citet{Bartlett_etal2017}, Bern10: \citet{Bernat_etal2010}, Bess91: \citet{Bessell1991}, Boes76: \citet{Boeshaar1976}, Bouy03: \citet{Bouy_etal2003}, Buen14: \citet{Buenzli_etal2014}, Burg99: \citet{Burgasser_etal1999}, Burg00a: \citet{Burgasser_etal2000}, Burg00b: \citet{Burgasser_etal2000b}, Burg02: \citet{Burgasser_etal2002}, Burg03a: \citet{Burgasser_etal2003c}, Burg03b: \citet{Burgasser_etal2003d}, Burg03c: \citet{Burgasser_etal2003b}, Burg03d: \citet{Burgasser_etal2003}, Burg04a: \citet{Burgasser2004c}, Burg04b: \citet{Burgasser_etal2004}, Burg05a: \citet{Burgasser_etal2005b}, Burg05b: \citet{Burgasser_etal2005}, Burg06a: \citet{Burgasser_etal2006}, Burg06b: \citet{Burgasser_etal2006b}, Burg06c: \citet{Burgasser_etal2006c}, Burg07: \citet{Burgasser_etal2007}, Burg08a: \citet{Burgasser_etal2008b}, Burg08b: \citet{Burgasser_etal2008c}, Burg10: \citet{Burgasser_etal2010b}, Burn08: \citet{Burningham_etal2008}, Burn09: \citet{Burningham_etal2009}, Chiu06: \citet{Chiu_etal2006}, Clar02: \citet{Clarke_etal2002}, Clar02b: \citet{Clarke_etal2002b}, Cruz03: \citet{Cruz_etal2003}, Cruz07: \citet{Cruz_etal2007}, Cruz09: \citet{Cruz_etal2009}, Dahn02: \citet{Dahn_etal2002}, Dahn08: \citet{Dahn_etal2008}, Delo08: \citet{Delorme_etal2008}, Diet14: \citet{Dieterich_etal2014}, Dupu12: \citet{Dupuy-Liu2012}, Dupu15: \citet{Dupuy_etal2015}, EROS99: \citet{EROSCollaboration1999}, Fan00: \citet{Fan_etal2000}, Folk07: \citet{Folkes_etal2007}, Gagn15: \citet{Gagne_etal2015}, Geba02: \citet{Geballe_etal2002}, Geli02: \citet{Gelino_etal2002}, Gira13: \citet{Girardin_etal2013}, Gizi97: \citet{Gizis1997}, Gizi00: \citet{Gizis_etal2000}, Gizi02: \citet{Gizis2002}, Gizi03: \citet{Gizis_etal2003}, Goli04: \citet{Golimowski_etal2004b}, Hard13: \citet{Harding_etal2013}, Hawl02: \citet{Hawley_etal2002}, Henr90: \citet{Henry-Kirkpatrick1990}, Irwi91: \citet{Irwin_etal1991}, Kao18: \citet{Kao_etal2018}, Kend04: \citet{Kendall_etal2004}, Kirk91: \citet{Kirkpatrick_etal1991}, Kirk95: \citet{Kirkpatrick_etal1995}, Kirk99: \citet{Kirkpatrick_etal1999}, Kirk00: \citet{Kirkpatrick_etal2000}, Kirk01: \citet{Kirkpatrick_etal2001}, Kirk08: \citet{Kirkpatrick_etal2008}, Kirk10: \citet{Kirkpatrick_etal2010}, Knap04: \citet{Knapp_etal2004}, Koen03: \citet{Koen2003}, Koen04: \citet{Koen_etal2004}, Koen13: \citet{Koen2013}, Legg96: \citet{Leggett_etal1996}, Legg00: \citet{Leggett_etal2000}, Lein86: \citet{Leinert_etal1986}, 
Lepi03: \citet{Lepine_etal2003}, Lieb03: \citet{Liebert_etal2003}, Liu05: \citet{Liu-Leggett2005}, Lodi05: \citet{Lodieu_etal2005}, Loop07: \citet{Looper_etal2007}, Loop08: \citet{Looper_etal2008b}, Luhm07a: \citet{Luhman_etal2007}, Luyt49: \citet{Luyten1949}, Mile17: \citet{Miles-Paez_etal2017b}, Maro13: \citet{Marocco_etal2013}, Maro15: \citet{Marocco_etal2015}, Mart99: \citet{Martin_etal1999}, Mart01: \citet{Martin_etal2001}, McCa04: \citet{McCaughrean_etal2004}, McEl06: \citet{McElwain-Burgasser2006}, Metc15: \citet{Metchev_etal2015}, Mora06: \citet{Morales-Calderon_etal2006}, Pine16: \citet{Pineda_etal2016}, Poko04: \citet{Pokorny_etal2004}, Pope13: \citet{Pope_etal2013}, Radi12: \citet{Radigan_etal2012}, Radi14: \citet{Radigan_etal2014}, Reid00: \citet{Reid_etal2000}, Reid01: \citet{Reid_etal2001}, Reid06: \citet{Reid_etal2006}, Reid08: \citet{Reid_etal2008}, Ruiz97: \citet{Ruiz_etal1997}, Schn14: \citet{Schneider_etal2014}, Scho03: \citet{Scholz_etal2003}, Scho04: \citet{Scholz_etal2004}, Sieg07: \citet{Siegler_etal2007}, Stra99: \citet{Strauss_etal1999}, Stum08: \citet{Stumpf_etal2008}, Thor03: \citet{Thorstensen-Kirkpatrick2003}, Tinn05: \citet{Tinney_etal2005}, vanB61: \citet{vanBiesbroeck1961}, Vos18: \citet{Vos_etal2018}, Vos20: \citet{Vos_etal2020}, Vos22: \citet{Vos_etal2022}, Warr07: \citet{Warren_etal2007}, Wils01: \citet{Wilson_etal2001}, Wils03: \citet{Wilson_etal2003b}, Wils14: \citet{Wilson_etal2014}, Zhou18: \citet{Zhou_etal2018}.\\
	(This table is available in machine-readable form in the online supplementary data.)\\
\end{tablenotes}
\end{ThreePartTable}
\end{landscape}
\clearpage
\twocolumn

\defcitealias{Burgasser_etal2008}{Burg08c}
\defcitealias{Cushing_etal2006}{Cush06}
\defcitealias{Filippazzo_etal2015}{Fili15}
\defcitealias{Kirkpatrick_etal2010}{Kirk10}
\defcitealias{Leggett_etal2009}{Legg09}
\defcitealias{Leggett_etal2010b}{Legg10b}
\defcitealias{Looper_etal2008b}{Loop08}
\defcitealias{Roellig_etal2004}{Roel04}
\defcitealias{Saumon_etal2007}{Saum07}
\defcitealias{Stephens_etal2009}{Step09}
\defcitealias{Suarez_etal2021a}{Suar21}

\clearpage
\onecolumn
\begin{landscape}
\begin{ThreePartTable}
\small
\begin{longtable}{@{\extracolsep{-8pt}}clllccccllcccc@{}}
	\caption{Log of the \textit{Spitzer} IRS observations of the 113 field M5--T9 dwarfs with detectable signal.} \label{tab:log} \\
	\toprule
	Dwarf           & Prog. ID & AOR KEY            & Modules       & \multicolumn{4}{c}{Exposure Time} & Obs. Date              & PI                  & \multicolumn{2}{c}{S/N at} & Publication             \\
	\cline{5-8}    
	\cline{11-12}    
	                &          &                    &               & SL2     & SL1 & LL2     & LL1     &                        &                     & 6 $\micron$ & 12 $\micron$ & Reference$^1$  \\
	(hhmm$\pm$ddmm) &          &                    &               & \multicolumn{4}{c}{(min)}         & (UT; YYYY/MM/DD)       &                     &             &              &                         \\
	\midrule
	\endfirsthead

	\multicolumn{14}{c}
	{{\bfseries \tablename\ \thetable{}} -- (Continued)} \\
	\toprule
	Dwarf           & Prog. ID & AOR KEY            & Modules       & \multicolumn{4}{c}{Exposure Time} & Obs. Date              & PI                  & \multicolumn{2}{c}{S/N at} & Publication             \\
	\cline{5-8}    
	\cline{11-12}    
	                &          &                    &               & SL2     & SL1 & LL2     & LL1     &                        &                     & 6 $\micron$ & 12 $\micron$ & Reference$^1$  \\
	(hhmm$\pm$ddmm) &          &                    &               & \multicolumn{4}{c}{(min)}         & (UT; YYYY/MM/DD)       &                     &             &              &                         \\
	\midrule
	\endhead

	0000+2554          & 20514    & 14796032, 14796288                     & SL2,SL1         & 144        & 200      & \nodata   & \nodata & 2005-12-18, 2006-01-25             & Golimowski, David A.         &  15         & 11     & \citetalias{Stephens_etal2009}            \\
	0004$-$4044        & 50367    & 26090752                               & SL2,SL1         & 2          & 3        & \nodata   & \nodata & 2009-01-15                         & Cushing, Michael C.          &  14         & 4      & Suar22                                    \\
	0024$-$0158        & 29       & 3877632                                & SL2,SL1,LL2     & 2          & 4        & 36        & \nodata & 2004-01-04                         & Houck, James R.              &  26         & 9      & \citetalias{Cushing_etal2006}             \\
	0025+4759          & 50367    & 26090496                               & SL2,SL1         & 16         & 30       & \nodata   & \nodata & 2009-03-07                         & Cushing, Michael C.          &  20         & 7      & \citetalias{Filippazzo_etal2015}, Suar22  \\
	0034$-$0052        & 40419    & 21830400, 21830656                     & SL1             & \nodata    & 512      & \nodata   & \nodata & 2008-01-15                         & Leggett, Sandy K.            &  \nodata    & 9      & \citetalias{Leggett_etal2009}             \\
	0034+0523          & 20544    & 14825728                               & SL2,SL1         & 120        & 120      & \nodata   & \nodata & 2006-01-19                         & Burgasser, Adam J.           &  9          & 12     & Suar22                                    \\
	0036+1821          & 51       & 4188672                                & SL2,SL1,LL2,LL1 & 8          & 8        & 8         & 16      & 2004-01-06                         & Houck, James R.              &  37         & 11     & \citetalias{Cushing_etal2006}             \\
\bottomrule
\end{longtable}
\begin{tablenotes}[para,flushleft]
	{\bf Note.} $^1$A second reference (Suar22) is added when the target spectrum is not shown nor available in the first reference.\\
	{\bf References.} Burg08c: \citet{Burgasser_etal2008}, Cush06: \citet{Cushing_etal2006}, Fili15: \citet{Filippazzo_etal2015}, Kirk10: \citet{Kirkpatrick_etal2010}, Legg09: \citet{Leggett_etal2009}, Legg10b: \citet{Leggett_etal2010b}, Loop08: \citet{Looper_etal2008b}, Roel04: \citet{Roellig_etal2004}, Saum07: \citet{Saumon_etal2007}, Suar21: \citet{Suarez_etal2021a}, Suar22: This work, Step09: \citet{Stephens_etal2009}. \\
	(This table is available in machine-readable form in the online supplementary data.)\\
\end{tablenotes}
\end{ThreePartTable}
\end{landscape}
\clearpage
\twocolumn

\defcitealias{Albert_etal2011}{Albe11}
\defcitealias{BardalezGagliuffi_etal2014}{Bard14}
\defcitealias{Becklin-Zuckerman1988}{Beck1988}
\defcitealias{Burgasser_etal2004}{Burg04b}
\defcitealias{Burgasser_etal2006b}{Burg06b}
\defcitealias{Cruz_etal2003}{Cruz03}
\defcitealias{Delfosse_etal1997}{Delf1997}
\defcitealias{Gagne_etal2015}{Gagn15}
\defcitealias{Geballe_etal2002}{Geba02}
\defcitealias{Hall2002}{Hall2002}
\defcitealias{Hawley_etal2002}{Hawl02}
\defcitealias{Kirkpatrick_etal1999}{Kirk99}
\defcitealias{Kirkpatrick_etal2008}{Kirk08}
\defcitealias{Lepine_etal2003b}{Lepi2003}
\defcitealias{Knapp_etal2004}{Knap04}
\defcitealias{Wilson_etal2003b}{Wils03}

\onecolumn
\begin{ThreePartTable}
\small
\begin{longtable}{@{\extracolsep{-8pt}}lcccccccc@{}}
	\caption{Select properties and observing log of the 8 field $\ge$M5 dwarfs with \textit{Spitzer} IRS observations that not yield detectable signal.} \label{tab:log_nullSNR} \\
	\toprule
	Designation             & Other Name     & R.A.$_{\textrm{J2000}}$ & Decl.$_{\textrm{J2000}}$ & Discovery Ref. & SpT$_{\textrm{opt}}$ & Ref.                       & SpT$_{\textrm{IR}}$ & Ref.                                    \\
	                        &                & ($^\circ$) & ($^\circ$) &                                    &                      &                                   &                     &                                         \\
	\midrule
	\endfirsthead

    2MASS J01075242+0041563 & \nodata        & 16.9680    & 0.6991     & \citetalias{Geballe_etal2002}      & L8                   & \citetalias{Hawley_etal2002}      & L7p                 & \citetalias{Gagne_etal2015}             \\
    2MASS J12095613-1004008 & \nodata        & 182.4840   & $-$10.0671 & \citetalias{Burgasser_etal2004}    & T3.5                 & \citetalias{Kirkpatrick_etal2008} & T3                  & \citetalias{Burgasser_etal2006b}        \\
    2MASS J12281523-1547342 & \nodata        & 187.0635   & $-$15.7930 & \citetalias{Delfosse_etal1997}     & L5                   & \citetalias{Kirkpatrick_etal1999} & L6                  & \citetalias{Knapp_etal2004}             \\
    2MASS J13153094-2649513 & \nodata        & 198.8786   & $-$26.8311 & \citetalias{Hall2002}              & L5.5                 & \citetalias{Kirkpatrick_etal2008} & L6.7                & \citetalias{BardalezGagliuffi_etal2014} \\
    2MASS J14243909+0917104 & GD 165B        & 216.1629   & 9.2863     & \citetalias{Becklin-Zuckerman1988} & L4                   & \citetalias{Kirkpatrick_etal1999} & L3                  & \citetalias{Knapp_etal2004}             \\
    2MASS J14250510+7102097 & LSR J1425+7102 & 216.2709   & 71.0360    & \citetalias{Lepine_etal2003b}      & sdM8                 & \citetalias{Lepine_etal2003b}     & \nodata             & \nodata                                 \\
    2MASS J14304358+2915405 & \nodata        & 217.6816   & 29.2613    & \citetalias{Wilson_etal2003b}      & L2                   & \citetalias{Cruz_etal2003}        & L1.5                & \citetalias{BardalezGagliuffi_etal2014} \\
    CFBDS J150000-182407    & \nodata        & 225.0014   & $-$18.4021 & \citetalias{Albert_etal2011}       & \nodata              & \nodata                           & T4.5                & \citetalias{Albert_etal2011}            \\
	\bottomrule
\end{longtable}
\end{ThreePartTable}

\addtocounter{table}{-1}
\begin{ThreePartTable}
\small
\begin{longtable}{@{\extracolsep{-8pt}}lllccccll@{}}
	\toprule
	Prog. ID & AOR KEY                    & Modules & \multicolumn{4}{c}{Exposure Time} & Obs. Date                          & PI                  \\
	\cline{4-7}    
	
	         &                            &         & SL2     & SL1 & LL2     & LL1     &                                    &                     \\
	         &                            &         & \multicolumn{4}{c}{(min)}         & (UT; YYYY/MM/DD)                   &                     \\
	\midrule
	\endfirsthead

    3136     & 12913920                   & SL2,SL1 & 8       & 8   & 0       & 0       & 2005-01-14                         & Cruz, Kelle         \\ 
    20514    & 14795264,14795520,14795776 & SL2,SL1 & 144     & 288 & 0       & 0       & 2006-01-30, 2006-01-31, 2006-02-01 & Golimowski, David A \\ 
    50367    & 26088448                   & SL2,SL1 & 12      & 24  & 0       & 0       & 2009-03-06                         & Cushing, Michael C  \\ 
    20716    & 15034112                   & SL2,SL1 & 16      & 16  & 0       & 0       & 2006-03-05                         & Gizis, John E       \\ 
    122      & 5020672                    & SL2,SL1 & 1       & 1   & 0       & 0       & 2004-06-24                         & Gehrz, Robert       \\ 
    251      & 12435968,13659136          & SL2,SL1 & 128     & 144 & 0       & 0       & 2005-02-10, 2005-06-04             & Burgasser, Adam J   \\ 
    50367    & 26087424                   & SL2,SL1 & 20      & 56  & 0       & 0       & 2009-03-06                         & Cushing, Michael C  \\ 
    50667    & 27414016                   & SL1     & 0       & 159 & 0       & 0       & 2009-04-01                         & Albert, Loic        \\ 
	\bottomrule
\end{longtable}
\begin{tablenotes}[para,flushleft]
	{\bf References.} Albe11: \citet{Albert_etal2011}, Bard14: \citet{BardalezGagliuffi_etal2014}, Beck1988: \citet{Becklin-Zuckerman1988}, Burg04b: \citet{Burgasser_etal2004}, Burg06b: \citet{Burgasser_etal2006b}, Cruz03: \citet{Cruz_etal2003}, Delf1997: \citet{Delfosse_etal1997}, Gagn15: \citet{Gagne_etal2015}, Geba02: \citet{Geballe_etal2002}, Hall2002: \citet{Hall2002}, Hawl02: \citet{Hawley_etal2002}, Kirk99: \citet{Kirkpatrick_etal1999}, Kirk08: \citet{Kirkpatrick_etal2008}, Lepi2003: \citet{Lepine_etal2003b}, Knap04: \citet{Knapp_etal2004}, Wils03: \citet{Wilson_etal2003b}. \\
\end{tablenotes}
\end{ThreePartTable}

\clearpage
\twocolumn

\defcitealias{Burgasser_etal2008}{Burg08c}
\defcitealias{Burningham_etal2008}{Burn08}
\defcitealias{Filippazzo_etal2015}{Fili15}
\defcitealias{Griffith_etal2012}{Grif12}
\defcitealias{Leggett_etal2007}{Legg07}
\defcitealias{Leggett_etal2010}{Legg10a}
\defcitealias{Luhman_etal2007}{Luhm07a}
\defcitealias{Luhman_etal2009}{Luhm09}
\defcitealias{Patten_etal2006}{Patt06}
\defcitealias{Warren_etal2007}{Warr07}

\clearpage
\onecolumn
\begin{ThreePartTable}
\small
\begin{longtable}{@{\extracolsep{-0pt}}cccccc@{}}
	\caption{Published IRAC [8.0] and WISE W3 photometry, and corresponding synthetic magnitudes of the 113 field M5--T9 dwarfs with IRS spectra.} \label{tab:photometry} \\
	\toprule
	Dwarf           & IRAC [8.0]       & IRAC Ref.        & WISE W3$^\textrm{a}$  &  IRAC [8.0]$_{\textrm{synt.}}^{\textrm{b}}$ & WISE W3$_{\textrm{synt.}}^{\textrm{b}}$ \\
	(hhmm$\pm$ddmm) & (mag)          &                  & (mag)                 &  (mag)                                    & (mag)                                    \\
	\midrule
	\endfirsthead

	\multicolumn{6}{c}
	{{\bfseries \tablename\ \thetable{}} -- (Continued)} \\
	\toprule
	Dwarf           & IRAC [8.0]       & IRAC Ref.        & WISE W3$^\textrm{a}$  &  IRAC [8.0]$_{\textrm{synt.}}^{\textrm{b}}$ & WISE W3$_{\textrm{synt.}}^{\textrm{b}}$ \\
	(hhmm$\pm$ddmm) & (mag)          &                  & (mag)                 &  (mag)                                    & (mag)                                    \\
	\midrule
	\endhead

	0000+2554    & 12.500$\pm$0.030   & \citetalias{Leggett_etal2007}    & \nodata          & 12.950$\pm$0.073 & \nodata         \\
	0004$-$4044  & 10.130$\pm$0.020   & \citetalias{Patten_etal2006}     &  9.803$\pm$0.048 & 10.413$\pm$0.136 & \nodata         \\
	0024$-$0158  &  9.550$\pm$0.010   & \citetalias{Patten_etal2006}     &  9.388$\pm$0.039 &  9.531$\pm$0.059 & 9.392$\pm$0.112 \\
	0025+4759    & \nodata            & \nodata                          & 11.364$\pm$0.121 & 11.459$\pm$0.089 & \nodata         \\
	0034$-$0052  & 13.910$\pm$0.060   & \citetalias{Warren_etal2007}     & 11.955$\pm$0.000 & \nodata          & \nodata         \\
	0034+0523    & 12.250$\pm$0.110   & \citetalias{Filippazzo_etal2015} & 11.780$\pm$0.313 & 12.387$\pm$0.116 & \nodata         \\
	0036+1821    & 10.060$\pm$0.010   & \citetalias{Patten_etal2006}     &  9.851$\pm$0.054 &  9.996$\pm$0.043 & 9.846$\pm$0.092 \\
	0059$-$0114  & \nodata            & \citetalias{Griffith_etal2012}   & 11.630$\pm$0.000 & \nodata          & \nodata         \\
	0136+0933    & \nodata            & \nodata                          &  9.738$\pm$0.048 & 10.287$\pm$0.058 & \nodata         \\
	0139$-$1757  & \nodata            & \nodata                          &  4.766$\pm$0.015 &  4.901$\pm$0.017 & 4.758$\pm$0.019 \\
	0144$-$0716  & \nodata            & \nodata                          & 10.883$\pm$0.100 & 11.063$\pm$0.089 & \nodata         \\
	0251$-$0352  & \nodata            & \nodata                          & 10.437$\pm$0.078 & 10.686$\pm$0.087 & \nodata         \\
	0255$-$4700  &  9.610$\pm$0.010   & \citetalias{Patten_etal2006}     &  9.142$\pm$0.027 &  9.621$\pm$0.027 & 9.176$\pm$0.035 \\
	0355+1133    & \nodata            & \nodata                          &  9.219$\pm$0.035 &  9.437$\pm$0.020 & \nodata         \\
	0415$-$0935  & 12.110$\pm$0.050   & \citetalias{Patten_etal2006}     & 11.093$\pm$0.125 & 12.010$\pm$0.044 & \nodata         \\
	0423$-$0414  & 11.010$\pm$0.030   & \citetalias{Patten_etal2006}     & 10.558$\pm$0.089 & 11.026$\pm$0.044 &10.619$\pm$0.089 \\
	0429$-$3123  & \nodata            & \nodata                          &  9.067$\pm$0.028 &  9.223$\pm$0.026 & \nodata         \\
	0439$-$2353  & 11.280$\pm$0.090   & \citetalias{Filippazzo_etal2015} & 11.074$\pm$0.121 & 11.268$\pm$0.079 & \nodata         \\
	0445$-$3048  & 10.980$\pm$0.070   & \citetalias{Filippazzo_etal2015} & 10.709$\pm$0.080 & 10.982$\pm$0.071 & \nodata         \\
	0501$-$0010  & 11.030$\pm$0.030   & \citetalias{Luhman_etal2009}     & 10.746$\pm$0.106 & 11.369$\pm$0.052 & \nodata         \\
	0523$-$1403  & 10.700$\pm$0.080   & \citetalias{Filippazzo_etal2015} & 10.375$\pm$0.071 & 10.697$\pm$0.075 & \nodata         \\
	0532+8246    & 13.030$\pm$0.100   & \citetalias{Patten_etal2006}     & 12.367$\pm$0.000 & 13.574$\pm$0.367 & \nodata         \\
	0539$-$0059  & 11.200$\pm$0.040   & \citetalias{Patten_etal2006}     & 11.716$\pm$0.288 & 11.204$\pm$0.069 &11.170$\pm$0.091 \\
	0559$-$1404  & 11.420$\pm$0.020   & \citetalias{Patten_etal2006}     & 10.605$\pm$0.089 & 11.387$\pm$0.076 &10.550$\pm$0.113 \\
	0624$-$4521  & 10.970$\pm$0.070   & \citetalias{Filippazzo_etal2015} & 10.760$\pm$0.072 & 10.968$\pm$0.060 & \nodata         \\
	0700+3157    &  9.980$\pm$0.060   & \citetalias{Filippazzo_etal2015} &  9.717$\pm$0.051 &  9.954$\pm$0.063 & \nodata         \\
	0727+1710    & 12.640$\pm$0.110   & \citetalias{Patten_etal2006}     & 11.924$\pm$0.309 & 12.639$\pm$0.108 & \nodata         \\
	0746+2000    &  9.570$\pm$0.010   & \citetalias{Patten_etal2006}     &  9.402$\pm$0.042 &  9.554$\pm$0.022 & 9.428$\pm$0.038 \\
	0758+3247    & 11.330$\pm$0.050   & \citetalias{Leggett_etal2007}    & 10.670$\pm$0.099 & 11.298$\pm$0.059 & \nodata         \\
	0805+4812    & 12.100$\pm$0.030   & \citetalias{Leggett_etal2007}    & 11.916$\pm$0.328 & 12.233$\pm$0.083 & \nodata         \\
	0825+2115    & 10.930$\pm$0.020   & \citetalias{Patten_etal2006}     & 10.509$\pm$0.096 & 10.928$\pm$0.072 &10.577$\pm$0.133 \\
	0829+2646    &  6.740$\pm$0.010   & \citetalias{Patten_etal2006}     &  6.618$\pm$0.016 &  6.712$\pm$0.028 & \nodata         \\
	0830+4828    & \nodata            & \nodata                          & 11.598$\pm$0.256 & 13.035$\pm$0.324 & \nodata         \\
	0837$-$0000  & 14.220$\pm$0.140   & \citetalias{Patten_etal2006}     & 12.712$\pm$0.000 & 13.859$\pm$0.418 & \nodata         \\
	0857+5708    & 10.740$\pm$0.020   & \citetalias{Patten_etal2006}     & 10.317$\pm$0.063 & 10.995$\pm$0.091 & \nodata         \\
	0908+5032    & 11.130$\pm$0.030   & \citetalias{Patten_etal2006}     & 10.748$\pm$0.114 & 11.247$\pm$0.076 & \nodata         \\
	0911+7401    & \nodata            & \nodata                          & 10.559$\pm$0.076 & 12.390$\pm$0.322 & \nodata         \\
	0912+1459    & 11.950$\pm$0.050   & \citetalias{Patten_etal2006}     & 11.297$\pm$0.189 & 11.954$\pm$0.082 & \nodata         \\
	0920+3517    & 12.210$\pm$0.220   & \citetalias{Filippazzo_etal2015} & 12.126$\pm$0.390 & 12.178$\pm$0.152 & \nodata         \\
	0921$-$2104  & \nodata            & \nodata                          & 10.387$\pm$0.069 & 10.621$\pm$0.079 & \nodata         \\
	0929+3429    & \nodata            & \nodata                          & 12.119$\pm$0.327 & 12.458$\pm$0.195 & \nodata         \\
	0937+2931    & 11.730$\pm$0.040   & \citetalias{Patten_etal2006}     & 10.696$\pm$0.101 & 11.653$\pm$0.065 & \nodata         \\
	0939$-$2448  & 11.890$\pm$0.030   & \citetalias{Burgasser_etal2008}  & 10.667$\pm$0.086 & 11.868$\pm$0.057 & \nodata         \\
	1013$-$1356  & \nodata            & \nodata                          & 12.359$\pm$0.000 & 13.476$\pm$0.218 & \nodata         \\
	1017+1308    & 11.700$\pm$0.030   & \citetalias{Patten_etal2006}     & 11.762$\pm$0.284 & 11.780$\pm$0.117 & \nodata         \\
	1021$-$0304  & 13.160$\pm$0.110   & \citetalias{Patten_etal2006}     & 12.286$\pm$0.424 & 13.313$\pm$0.229 & \nodata         \\
	1022+5825    & \nodata            & \nodata                          & 11.156$\pm$0.137 & 13.071$\pm$0.434 & \nodata         \\
	1028+5654    & \nodata            & \nodata                          & \nodata          & \nodata          & \nodata         \\
	1036$-$3441  & 11.890$\pm$0.140   & \citetalias{Filippazzo_etal2015} & 11.454$\pm$0.185 & 11.860$\pm$0.061 & \nodata         \\
	1045$-$0149  & \nodata            & \nodata                          & 10.768$\pm$0.109 & 10.938$\pm$0.098 & \nodata         \\
	1048+0111    & \nodata            & \nodata                          & 10.582$\pm$0.111 & 10.754$\pm$0.086 & \nodata         \\
	1051+5613    & \nodata            & \nodata                          & 10.782$\pm$0.095 & 10.981$\pm$0.093 & \nodata         \\
	1052+4422    & 12.360$\pm$0.030   & \citetalias{Leggett_etal2007}    & 12.530$\pm$0.420 & 12.428$\pm$0.069 & \nodata         \\
	1106+2754    & \nodata            & \nodata                          & 11.526$\pm$0.195 & 11.784$\pm$0.063 & \nodata         \\
	1108+6830    & \nodata            & \nodata                          & 10.168$\pm$0.049 & 10.348$\pm$0.031 &10.175$\pm$0.058 \\
	1110+0116    & 13.210$\pm$0.160   & \citetalias{Patten_etal2006}     & 12.000$\pm$0.294 & 13.061$\pm$0.108 & \nodata         \\
	1112+3548    & \nodata            & \nodata                          & 11.036$\pm$0.133 & 11.319$\pm$0.110 & \nodata         \\
	1114$-$2618  & 12.250$\pm$0.040   & \citetalias{Leggett_etal2010}    & 11.346$\pm$0.179 & 12.261$\pm$0.078 & \nodata         \\
	1126$-$5003  & \nodata            & \nodata                          & 11.584$\pm$0.146 & 11.619$\pm$0.052 & \nodata         \\
	1155+0559    & 12.260$\pm$0.040   & \citetalias{Leggett_etal2007}    & 11.968$\pm$0.331 & 12.163$\pm$0.090 & \nodata         \\
	1207+0244    & 12.230$\pm$0.060   & \citetalias{Leggett_etal2007}    & 11.467$\pm$0.214 & 12.125$\pm$0.064 & \nodata         \\
	1213$-$0432  & \nodata            & \nodata                          & 11.052$\pm$0.216 & 11.822$\pm$0.155 & \nodata         \\
	1217$-$0311  & 12.950$\pm$0.180   & \citetalias{Patten_etal2006}     & 11.704$\pm$0.264 & 12.701$\pm$0.326 & \nodata         \\
	1225$-$2739  & 12.240$\pm$0.020   & \citetalias{Patten_etal2006}     & 11.157$\pm$0.146 & 12.307$\pm$0.201 &11.255$\pm$0.219 \\
	1237+6526    & 12.780$\pm$0.110   & \citetalias{Patten_etal2006}     & 11.939$\pm$0.224 & 13.000$\pm$0.218 & \nodata         \\
	1239+5515    & 11.430$\pm$0.040   & \citetalias{Filippazzo_etal2015} & 11.108$\pm$0.095 & 11.357$\pm$0.067 &11.054$\pm$0.116 \\
	1254$-$0122  & 11.750$\pm$0.040   & \citetalias{Patten_etal2006}     & 10.972$\pm$0.137 & 11.743$\pm$0.103 &11.026$\pm$0.181 \\
	1305$-$2541  & 10.610$\pm$0.020   & \citetalias{Patten_etal2006}     & 10.418$\pm$0.064 & 10.559$\pm$0.046 & \nodata         \\
	1331$-$0116  & 12.620$\pm$0.060   & \citetalias{Patten_etal2006}     & 12.489$\pm$0.435 & 12.506$\pm$0.098 & \nodata         \\
	1335+1130    & 13.370$\pm$0.070   & \citetalias{Burningham_etal2008} & 12.151$\pm$0.358 & \nodata          & \nodata         \\
	1425$-$3650  & 10.110$\pm$0.050   & \citetalias{Filippazzo_etal2015} & 10.019$\pm$0.047 & 10.122$\pm$0.059 & \nodata         \\
	1439+1839    & \nodata            & \nodata                          & 11.859$\pm$0.203 & 11.909$\pm$0.054 & \nodata         \\
	1439+1929    & 10.670$\pm$0.020   & \citetalias{Patten_etal2006}     & 10.523$\pm$0.057 & 10.635$\pm$0.046 & \nodata         \\
	1441$-$0945  & \nodata            & \nodata                          & 12.110$\pm$0.338 & 11.809$\pm$0.128 & \nodata         \\
	1448+1031    & 11.060$\pm$0.080   & \citetalias{Filippazzo_etal2015} & 10.827$\pm$0.088 & 11.077$\pm$0.083 & \nodata         \\
	1456$-$2809  &  8.360$\pm$0.010   & \citetalias{Patten_etal2006}     &  8.273$\pm$0.025 &  8.336$\pm$0.050 & 8.254$\pm$0.071 \\
	1457$-$2121  & 11.970$\pm$0.070   & \citetalias{Patten_etal2006}     & 10.784$\pm$0.086 & 11.994$\pm$0.099 & \nodata         \\
	1503+2525    & 11.540$\pm$0.040   & \citetalias{Griffith_etal2012}   & 10.709$\pm$0.071 & 11.422$\pm$0.080 & \nodata         \\
	1506+1321    & 10.580$\pm$0.010   & \citetalias{Patten_etal2006}     & 10.491$\pm$0.062 & 10.547$\pm$0.046 & \nodata         \\
	1507$-$1627  &  9.990$\pm$0.010   & \citetalias{Patten_etal2006}     &  9.685$\pm$0.048 &  9.976$\pm$0.035 & \nodata         \\
	1515+4847    & 10.830$\pm$0.020   & \citetalias{Patten_etal2006}     & 10.461$\pm$0.050 & 10.888$\pm$0.050 & \nodata         \\
	1516+3053    & 12.470$\pm$0.040   & \citetalias{Leggett_etal2007}    & 12.093$\pm$0.203 & 12.513$\pm$0.064 & \nodata         \\
	1520+3546    & 12.260$\pm$0.030   & \citetalias{Leggett_etal2007}    & 11.998$\pm$0.186 & 12.202$\pm$0.066 & \nodata         \\
	1523+3014    & 12.210$\pm$0.140   & \citetalias{Filippazzo_etal2015} & 11.905$\pm$0.181 & 12.279$\pm$0.088 & \nodata         \\
	1526+2043    & 12.320$\pm$0.040   & \citetalias{Patten_etal2006}     & 12.301$\pm$0.256 & 12.272$\pm$0.144 & \nodata         \\
	1610$-$0040  & \nodata            & \nodata                          & 11.088$\pm$0.134 & 11.349$\pm$0.087 & \nodata         \\
	1615+3559    & \nodata            & \nodata                          & 11.434$\pm$0.132 & 11.686$\pm$0.121 & \nodata         \\
	1624+0029    & 12.840$\pm$0.090   & \citetalias{Patten_etal2006}     & 12.163$\pm$0.394 & 12.789$\pm$0.145 & \nodata         \\
	1626+3925    & \nodata            & \nodata                          & 12.174$\pm$0.000 & 12.826$\pm$0.115 & \nodata         \\
	1658+7027    & \nodata            & \nodata                          & 10.844$\pm$0.058 & 11.084$\pm$0.052 &10.854$\pm$0.108 \\
	1707$-$0558  & \nodata            & \nodata                          &  9.399$\pm$0.045 &  9.586$\pm$0.036 & \nodata         \\
	1721+3344    & 11.400$\pm$0.020   & \citetalias{Patten_etal2006}     & 11.123$\pm$0.110 & 11.452$\pm$0.038 & \nodata         \\
	1728+3948    & 12.130$\pm$0.030   & \citetalias{Patten_etal2006}     & 11.743$\pm$0.150 & 12.156$\pm$0.121 & \nodata         \\
	1731+2721    & \nodata            & \nodata                          &  9.875$\pm$0.043 &  9.996$\pm$0.039 & \nodata         \\
	1753$-$6559  & \nodata            & \nodata                          & 11.032$\pm$0.099 & 11.406$\pm$0.115 & \nodata         \\
	1807+5015    & \nodata            & \nodata                          & 10.505$\pm$0.056 & 10.667$\pm$0.068 & \nodata         \\
	1821+1414    & \nodata            & \nodata                          &  9.928$\pm$0.052 & 10.097$\pm$0.023 & \nodata         \\
	1828+1229    & \nodata            & \nodata                          & 12.683$\pm$0.505 & 13.052$\pm$0.269 & \nodata         \\
	1916+0509    & 8.140$\pm$0.000    & \citetalias{Patten_etal2006}     &  8.041$\pm$0.021 &  8.111$\pm$0.033 & 8.053$\pm$0.050 \\
	1936$-$5502  & \nodata            & \nodata                          & 11.682$\pm$0.384 & 11.748$\pm$0.123 & \nodata         \\
	2057$-$0252  & \nodata            & \nodata                          & 10.277$\pm$0.080 & 10.641$\pm$0.086 & \nodata         \\
	2132+1341    & \nodata            & \nodata                          & 11.645$\pm$0.000 & 12.487$\pm$0.100 & \nodata         \\
	2139+0220    & \nodata            & \nodata                          & 10.496$\pm$0.084 & 11.657$\pm$0.200 & \nodata         \\
	2144+1446    & 12.580$\pm$0.110   & \citetalias{Luhman_etal2007}     & 11.777$\pm$0.217 & 12.526$\pm$0.042 & \nodata         \\
	2146$-$0010  & 14.360$\pm$0.080   & \citetalias{Leggett_etal2010}    & 12.312$\pm$0.000 & \nodata          & \nodata         \\
	2148+4003    & \nodata            & \nodata                          &  9.631$\pm$0.042 &  9.841$\pm$0.019 & \nodata         \\
	2152+0937    & \nodata            & \nodata                          & 11.207$\pm$0.172 & 11.544$\pm$0.098 & \nodata         \\
	2204$-$5646  &  8.980$\pm$0.040   & \citetalias{Patten_etal2006}     &  8.393$\pm$0.023 &  9.023$\pm$0.021 & 8.275$\pm$0.025 \\
	2224$-$0158  & 10.810$\pm$0.020   & \citetalias{Patten_etal2006}     & 10.540$\pm$0.101 & 10.861$\pm$0.061 & \nodata         \\
	2238$-$1517  & \nodata            & \nodata                          &  5.005$\pm$0.015 &  5.306$\pm$0.018 & 5.291$\pm$0.022 \\
	2244+2043    & 11.360$\pm$0.030   & \citetalias{Leggett_etal2007}    & 11.225$\pm$0.141 & 11.387$\pm$0.059 & \nodata         \\
	2254+3123    & 12.780$\pm$0.100   & \citetalias{Patten_etal2006}     & 12.111$\pm$0.340 & 12.823$\pm$0.065 & \nodata         \\
	2351$-$2537  & \nodata            & \nodata                          & 10.262$\pm$0.073 & 10.523$\pm$0.072 & \nodata         \\
	\bottomrule
\end{longtable}
\begin{tablenotes}[para,flushleft]
	{\bf Notes.}\\
	$^\textrm{a}$ WISE W3 photometry from \citet{Cutri_etal2013}. \\
	$^\textrm{b}$ Synthetic magnitudes (corrected by the $-0.064$ mag mean offset with respect to the observations) from the IRS spectra.\\
	{\bf References.} Burg08c: \citet{Burgasser_etal2008}, Fili15: \citet{Filippazzo_etal2015}, Grif12: \citet{Griffith_etal2012}, Legg07: \citet{Leggett_etal2007}, Legg10a: \citet{Leggett_etal2010}, Luhm07a: \citet{Luhman_etal2007}, Luhm09: \citet{Luhman_etal2009}, Patt06: \citet{Patten_etal2006}, Warr07: \citet{Warren_etal2007}. \\
	(This table is available in machine-readable form in the online supplementary data.)\\
\end{tablenotes}
\end{ThreePartTable}
\clearpage
\twocolumn

\clearpage
\onecolumn
\begin{ThreePartTable}
\small
\begin{longtable}{@{\extracolsep{-0pt}}ccccc@{}}
	\caption{Spectral index values of the principal mid-infrared absorption bands for the 113 field M5--T9 dwarfs with IRS spectra.} \label{tab:indices} \\
	\toprule
    Dwarf           & H$_2$O Index & CH$_4$ Index & NH$_3$ Index & Silicate Index \\
	(hhmm$\pm$ddmm) &              &              &              &                \\
	\midrule
	\endfirsthead

	\multicolumn{5}{c}
	{{\bfseries \tablename\ \thetable{}} -- (Continued)} \\
	\toprule
    Dwarf           & H$_2$O Index & CH$_4$ Index & NH$_3$ Index & Silicate Index \\
	(hhmm$\pm$ddmm) &              &              &              &                \\
	\midrule
	\endhead

	0000+2554    &  1.44$\pm$0.12     & 2.92$\pm$0.25     & 1.24$\pm$0.05    &  0.72$\pm$0.14      \\
	0004$-$4044  &  1.12$\pm$0.02     & 0.67$\pm$0.05     & 1.05$\pm$0.07    &  1.15$\pm$0.21      \\
	0024$-$0158  &  1.03$\pm$0.01     & 0.67$\pm$0.01     & 1.10$\pm$0.02    &  1.01$\pm$0.04      \\
	0025+4759    &  1.07$\pm$0.01     & 0.63$\pm$0.03     & 1.03$\pm$0.05    &  1.35$\pm$0.10      \\
	0034$-$0052  &  \nodata           & 4.59$\pm$0.48     & 1.63$\pm$0.18    &  \nodata            \\
	0034+0523    &  1.32$\pm$0.09     & 2.47$\pm$0.12     & 1.36$\pm$0.05    &  0.74$\pm$0.12      \\
	0036+1821    &  1.12$\pm$0.01     & 0.59$\pm$0.01     & 1.02$\pm$0.03    &  1.28$\pm$0.08      \\
	0059$-$0114  &  \nodata           & 3.52$\pm$0.37     & 1.60$\pm$0.16    &  \nodata            \\
	0136+0933    &  1.20$\pm$0.03     & 1.56$\pm$0.04     & 1.21$\pm$0.02    &  0.71$\pm$0.08      \\
	0139$-$1757  &  1.05$\pm$0.01     & 0.67$\pm$0.01     & 1.13$\pm$0.01    &  1.09$\pm$0.02      \\
	0144$-$0716  &  1.05$\pm$0.02     & 0.57$\pm$0.02     & 1.00$\pm$0.04    &  1.46$\pm$0.06      \\
	0251$-$0352  &  1.11$\pm$0.02     & 0.56$\pm$0.02     & 1.13$\pm$0.04    &  1.16$\pm$0.06      \\
	0255$-$4700  &  1.33$\pm$0.03     & 1.03$\pm$0.02     & 1.09$\pm$0.02    &  0.84$\pm$0.05      \\
	0355+1133    &  1.00$\pm$0.01     & 0.55$\pm$0.01     & 1.05$\pm$0.01    &  1.35$\pm$0.03      \\
	0415$-$0935  &  1.58$\pm$0.14     & 3.85$\pm$0.23     & 1.56$\pm$0.11    &  0.61$\pm$0.09      \\
	0423$-$0414  &  1.27$\pm$0.03     & 1.03$\pm$0.02     & 1.10$\pm$0.02    &  0.88$\pm$0.05      \\
	0429$-$3123  &  1.04$\pm$0.01     & 0.67$\pm$0.01     & 1.16$\pm$0.02    &  1.09$\pm$0.04      \\
	0439$-$2353  &  1.24$\pm$0.03     & 0.71$\pm$0.02     & 1.06$\pm$0.03    &  1.12$\pm$0.08      \\
	0445$-$3048  &  1.04$\pm$0.02     & 0.70$\pm$0.02     & 1.15$\pm$0.04    &  1.10$\pm$0.07      \\
	0501$-$0010  &  1.01$\pm$0.02     & 0.46$\pm$0.03     & 1.04$\pm$0.04    &  1.45$\pm$0.19      \\
	0523$-$1403  &  1.13$\pm$0.01     & 0.61$\pm$0.03     & 1.14$\pm$0.06    &  1.19$\pm$0.11      \\
	0532+8246    &  1.18$\pm$0.05     & 0.47$\pm$0.06     & 1.09$\pm$0.17    &  1.08$\pm$0.21      \\
	0539$-$0059  &  1.35$\pm$0.03     & 0.67$\pm$0.04     & 0.94$\pm$0.05    &  1.05$\pm$0.19      \\
	0559$-$1404  &  1.17$\pm$0.05     & 2.10$\pm$0.10     & 1.32$\pm$0.04    &  0.71$\pm$0.08      \\
	0624$-$4521  &  1.12$\pm$0.02     & 0.68$\pm$0.01     & 1.05$\pm$0.03    &  1.23$\pm$0.06      \\
	0700+3157    &  1.16$\pm$0.01     & 0.70$\pm$0.01     & 1.10$\pm$0.03    &  1.05$\pm$0.05      \\
	0727+1710    &  1.12$\pm$0.06     & 2.67$\pm$0.16     & 1.51$\pm$0.06    &  0.66$\pm$0.13      \\
	0746+2000    &  1.08$\pm$0.02     & 0.65$\pm$0.01     & 1.10$\pm$0.02    &  1.09$\pm$0.03      \\
	0758+3247    &  1.17$\pm$0.07     & 1.09$\pm$0.08     & 0.98$\pm$0.04    &  0.74$\pm$0.15      \\
	0805+4812    &  1.14$\pm$0.06     & 0.84$\pm$0.03     & 1.02$\pm$0.05    &  1.19$\pm$0.13      \\
	0825+2115    &  1.19$\pm$0.02     & 0.69$\pm$0.02     & 1.05$\pm$0.03    &  1.05$\pm$0.09      \\
	0829+2646    &  1.07$\pm$0.01     & 0.70$\pm$0.01     & 1.17$\pm$0.02    &  1.05$\pm$0.03      \\
	0830+4828    &  1.38$\pm$0.13     & 1.57$\pm$0.20     & 1.27$\pm$0.13    &  0.61$\pm$0.21      \\
	0837$-$0000  &  1.34$\pm$0.11     & 1.24$\pm$0.13     & 1.07$\pm$0.10    &  0.82$\pm$0.21      \\
	0857+5708    &  1.26$\pm$0.03     & 0.72$\pm$0.02     & 1.12$\pm$0.02    &  1.02$\pm$0.09      \\
	0908+5032    &  1.23$\pm$0.06     & 0.81$\pm$0.04     & 1.14$\pm$0.05    &  0.98$\pm$0.09      \\
	0911+7401    &  1.10$\pm$0.05     & 0.78$\pm$0.08     & 1.21$\pm$0.13    &  1.21$\pm$0.32      \\
	0912+1459    &  1.33$\pm$0.04     & 1.29$\pm$0.05     & 1.22$\pm$0.05    &  0.87$\pm$0.10      \\
	0920+3517    &  1.24$\pm$0.07     & 0.68$\pm$0.05     & 0.95$\pm$0.07    &  1.33$\pm$0.25      \\
	0921$-$2104  &  1.11$\pm$0.01     & 0.70$\pm$0.02     & 1.15$\pm$0.03    &  1.00$\pm$0.07      \\
	0929+3429    &  1.17$\pm$0.11     & 0.70$\pm$0.13     & 0.76$\pm$0.16    &  1.24$\pm$0.31      \\
	0937+2931    &  1.33$\pm$0.08     & 2.52$\pm$0.09     & 1.52$\pm$0.06    &  0.60$\pm$0.08      \\
	0939$-$2448  &  1.66$\pm$0.15     & 4.93$\pm$0.24     & 1.58$\pm$0.12    &  0.60$\pm$0.11      \\
	1013$-$1356  &  1.11$\pm$0.07     & 0.63$\pm$0.07     & 0.92$\pm$0.11    &  1.52$\pm$0.34      \\
	1017+1308    &  1.01$\pm$0.03     & 0.74$\pm$0.04     & 1.10$\pm$0.04    &  1.07$\pm$0.13      \\
	1021$-$0304  &  1.19$\pm$0.07     & 1.77$\pm$0.18     & 1.20$\pm$0.07    &  0.89$\pm$0.18      \\
	1022+5825    &  1.23$\pm$0.09     & 0.74$\pm$0.12     & 0.82$\pm$0.12    &  1.27$\pm$0.42      \\
	1028+5654    &  \nodata           & 2.92$\pm$0.41     & 1.24$\pm$0.10    &  \nodata            \\
	1036$-$3441  &  1.32$\pm$0.04     & 0.98$\pm$0.03     & 1.11$\pm$0.03    &  0.88$\pm$0.08      \\
	1045$-$0149  &  1.07$\pm$0.02     & 0.74$\pm$0.02     & 1.10$\pm$0.03    &  0.98$\pm$0.08      \\
	1048+0111    &  1.06$\pm$0.02     & 0.73$\pm$0.02     & 1.09$\pm$0.04    &  0.99$\pm$0.07      \\
	1051+5613    &  1.08$\pm$0.02     & 0.70$\pm$0.02     & 1.36$\pm$0.07    &  1.12$\pm$0.10      \\
	1052+4422    &  1.26$\pm$0.06     & 0.79$\pm$0.06     & 1.05$\pm$0.04    &  0.97$\pm$0.16      \\
	1106+2754    &  1.28$\pm$0.04     & 1.39$\pm$0.04     & 1.23$\pm$0.03    &  0.76$\pm$0.05      \\
	1108+6830    &  1.02$\pm$0.02     & 0.65$\pm$0.02     & 1.11$\pm$0.03    &  1.07$\pm$0.07      \\
	1110+0116    &  1.19$\pm$0.05     & 2.36$\pm$0.16     & 1.14$\pm$0.06    &  0.62$\pm$0.09      \\
	1112+3548    &  1.15$\pm$0.02     & 0.68$\pm$0.02     & 0.98$\pm$0.05    &  1.37$\pm$0.11      \\
	1114$-$2618  &  1.39$\pm$0.16     & 3.40$\pm$0.26     & 1.52$\pm$0.09    &  0.60$\pm$0.11      \\
	1126$-$5003  &  1.14$\pm$0.02     & 0.69$\pm$0.02     & 1.15$\pm$0.03    &  1.12$\pm$0.08      \\
	1155+0559    &  0.83$\pm$0.04     & 0.73$\pm$0.05     & 1.34$\pm$0.06    &  1.06$\pm$0.16      \\
	1207+0244    &  1.18$\pm$0.05     & 0.94$\pm$0.03     & 1.06$\pm$0.02    &  0.86$\pm$0.09      \\
	1213$-$0432  &  1.18$\pm$0.03     & 0.65$\pm$0.04     & 1.00$\pm$0.07    &  1.16$\pm$0.16      \\
	1217$-$0311  &  1.28$\pm$0.12     & 3.03$\pm$0.24     & 1.71$\pm$0.13    &  0.77$\pm$0.15      \\
	1225$-$2739  &  1.32$\pm$0.10     & 2.66$\pm$0.18     & 1.55$\pm$0.08    &  0.65$\pm$0.09      \\
	1237+6526    &  1.19$\pm$0.10     & 1.61$\pm$0.26     & 1.30$\pm$0.14    &  0.91$\pm$0.25      \\
	1239+5515    &  1.20$\pm$0.03     & 0.63$\pm$0.03     & 1.17$\pm$0.06    &  1.68$\pm$0.22      \\
	1254$-$0122  &  1.29$\pm$0.06     & 1.75$\pm$0.11     & 1.23$\pm$0.05    &  0.63$\pm$0.11      \\
	1305$-$2541  &  1.08$\pm$0.01     & 0.61$\pm$0.01     & 1.09$\pm$0.03    &  1.15$\pm$0.05      \\
	1331$-$0116  &  1.25$\pm$0.08     & 0.73$\pm$0.03     & 1.02$\pm$0.03    &  1.02$\pm$0.13      \\
	1335+1130    &  \nodata           & 6.00$\pm$1.12     & 1.56$\pm$0.13    &  \nodata            \\
	1425$-$3650  &  1.04$\pm$0.01     & 0.55$\pm$0.01     & 1.08$\pm$0.03    &  1.36$\pm$0.06      \\
	1439+1839    &  1.06$\pm$0.03     & 0.66$\pm$0.02     & 1.05$\pm$0.04    &  1.11$\pm$0.11      \\
	1439+1929    &  1.09$\pm$0.01     & 0.74$\pm$0.02     & 1.14$\pm$0.03    &  1.04$\pm$0.05      \\
	1441$-$0945  &  1.15$\pm$0.05     & 0.84$\pm$0.06     & 1.29$\pm$0.11    &  0.97$\pm$0.20      \\
	1448+1031    &  1.15$\pm$0.01     & 0.58$\pm$0.02     & 0.96$\pm$0.03    &  1.37$\pm$0.10      \\
	1456$-$2809  &  1.06$\pm$0.01     & 0.69$\pm$0.02     & 1.19$\pm$0.02    &  1.10$\pm$0.04      \\
	1457$-$2121  &  1.32$\pm$0.10     & 4.29$\pm$0.18     & 1.52$\pm$0.07    &  0.59$\pm$0.09      \\
	1503+2525    &  1.24$\pm$0.05     & 2.51$\pm$0.11     & 1.47$\pm$0.05    &  0.68$\pm$0.07      \\
	1506+1321    &  1.07$\pm$0.01     & 0.56$\pm$0.01     & 1.08$\pm$0.02    &  1.25$\pm$0.05      \\
	1507$-$1627  &  1.24$\pm$0.02     & 0.73$\pm$0.01     & 1.07$\pm$0.02    &  1.12$\pm$0.04      \\
	1515+4847    &  1.28$\pm$0.02     & 0.80$\pm$0.02     & 1.09$\pm$0.02    &  0.95$\pm$0.06      \\
	1516+3053    &  1.37$\pm$0.11     & 0.98$\pm$0.06     & 1.19$\pm$0.05    &  0.76$\pm$0.11      \\
	1520+3546    &  1.30$\pm$0.09     & 1.09$\pm$0.06     & 1.17$\pm$0.04    &  0.78$\pm$0.09      \\
	1523+3014    &  1.21$\pm$0.03     & 1.01$\pm$0.04     & 1.13$\pm$0.04    &  0.85$\pm$0.09      \\
	1526+2043    &  1.29$\pm$0.08     & 0.84$\pm$0.07     & 1.08$\pm$0.10    &  0.92$\pm$0.19      \\
	1610$-$0040  &  1.01$\pm$0.01     & 0.68$\pm$0.03     & 1.05$\pm$0.04    &  1.15$\pm$0.08      \\
	1615+3559    &  1.09$\pm$0.02     & 0.55$\pm$0.02     & 1.19$\pm$0.07    &  1.30$\pm$0.13      \\
	1624+0029    &  1.24$\pm$0.09     & 2.44$\pm$0.12     & 1.43$\pm$0.06    &  0.62$\pm$0.08      \\
	1626+3925    &  0.95$\pm$0.05     & 0.69$\pm$0.05     & 1.08$\pm$0.08    &  1.08$\pm$0.13      \\
	1658+7027    &  1.09$\pm$0.02     & 0.74$\pm$0.02     & 1.04$\pm$0.04    &  0.97$\pm$0.07      \\
	1707$-$0558  &  1.08$\pm$0.01     & 0.70$\pm$0.01     & 1.10$\pm$0.02    &  1.10$\pm$0.04      \\
	1721+3344    &  1.10$\pm$0.02     & 0.62$\pm$0.02     & 1.11$\pm$0.03    &  1.19$\pm$0.09      \\
	1728+3948    &  1.09$\pm$0.05     & 0.74$\pm$0.05     & 1.06$\pm$0.08    &  1.22$\pm$0.18      \\
	1731+2721    &  1.04$\pm$0.01     & 0.72$\pm$0.01     & 1.17$\pm$0.02    &  1.00$\pm$0.04      \\
	1753$-$6559  &  1.06$\pm$0.02     & 0.50$\pm$0.02     & 0.91$\pm$0.04    &  1.51$\pm$0.18      \\
	1807+5015    &  1.05$\pm$0.01     & 0.65$\pm$0.02     & 1.13$\pm$0.03    &  1.16$\pm$0.07      \\
	1821+1414    &  1.16$\pm$0.01     & 0.61$\pm$0.01     & 1.00$\pm$0.01    &  1.38$\pm$0.03      \\
	1828+1229    &  1.09$\pm$0.05     & 0.70$\pm$0.06     & 1.10$\pm$0.11    &  1.05$\pm$0.16      \\
	1916+0509    &  1.08$\pm$0.01     & 0.68$\pm$0.01     & 1.16$\pm$0.03    &  1.06$\pm$0.02      \\
	1936$-$5502  &  1.21$\pm$0.03     & 0.70$\pm$0.04     & 1.09$\pm$0.08    &  1.21$\pm$0.12      \\
	2057$-$0252  &  1.06$\pm$0.02     & 0.69$\pm$0.03     & 1.11$\pm$0.05    &  1.05$\pm$0.12      \\
	2132+1341    &  1.15$\pm$0.05     & 0.66$\pm$0.04     & 1.04$\pm$0.05    &  1.27$\pm$0.14      \\
	2139+0220    &  1.22$\pm$0.04     & 0.99$\pm$0.07     & 0.99$\pm$0.06    &  0.72$\pm$0.13      \\
	2144+1446    &  1.18$\pm$0.04     & 1.65$\pm$0.06     & 1.24$\pm$0.03    &  0.69$\pm$0.06      \\
	2146$-$0010  &  \nodata           & 5.53$\pm$0.85     & 1.38$\pm$0.12    &  \nodata            \\
	2148+4003    &  1.08$\pm$0.01     & 0.57$\pm$0.01     & 0.97$\pm$0.01    &  1.58$\pm$0.04      \\
	2152+0937    &  1.29$\pm$0.03     & 0.73$\pm$0.03     & 1.11$\pm$0.04    &  1.10$\pm$0.08      \\
	2204$-$5646  &  1.32$\pm$0.04     & 1.89$\pm$0.05     & 1.32$\pm$0.03    &  0.67$\pm$0.05      \\
	2224$-$0158  &  1.10$\pm$0.02     & 0.57$\pm$0.02     & 0.96$\pm$0.02    &  1.48$\pm$0.11      \\
	2238$-$1517  &  1.04$\pm$0.01     & 0.60$\pm$0.02     & 1.16$\pm$0.02    &  1.16$\pm$0.06      \\
	2244+2043    &  1.15$\pm$0.03     & 0.49$\pm$0.02     & 0.92$\pm$0.04    &  1.42$\pm$0.15      \\
	2254+3123    &  1.30$\pm$0.06     & 1.64$\pm$0.08     & 1.26$\pm$0.05    &  0.82$\pm$0.11      \\
	2351$-$2537  &  1.08$\pm$0.01     & 0.69$\pm$0.02     & 1.23$\pm$0.03    &  1.06$\pm$0.07      \\
	\bottomrule
\end{longtable}
\begin{tablenotes}[para,flushleft]
	(This table is available in machine-readable form in the online supplementary data.)\\
\end{tablenotes}
\begin{tablenotes}[para,flushleft]
\end{tablenotes}
\end{ThreePartTable}
\clearpage
\twocolumn

\bibliographystyle{mnras}
\bibliography{mybib_Suarez} 





\bsp	
\label{lastpage}
\end{document}